\documentclass[fleqn,usenatbib]{mnras}

\usepackage{newtxtext,newtxmath}
\usepackage[T1]{fontenc}

\DeclareRobustCommand{\VAN}[3]{#2}
\let\VANthebibliography\thebibliography
\def\thebibliography{\DeclareRobustCommand{\VAN}[3]{##3}\VANthebibliography}

\usepackage{graphicx}	% Including figure files
\usepackage{amsmath}	% Advanced maths commands
\usepackage{color}
\usepackage{float}
\usepackage{pdflscape}
\usepackage{footnote}
\usepackage{rotating}
\usepackage{bm}
\usepackage{threeparttable}
\newcommand{\cxo}{{\it Chandra}}

\newcommand{\ergcm}[1]{erg\,cm$^{-2}$\,s$^{-1}$}

\def\HI{\hbox{H{\sc i}}}
\def\HII{\hbox{H{\sc ii}}}

\newcommand{\OIII}{[O\,{\sc iii}]}

\newcommand{\FeXIV}{[Fe\,{\sc xiv}]}
\newcommand{\FeXI}{[Fe\,{\sc xi}]}

\newcommand{\kms}{km\,s$^{-1}$}

\title[ATCA Study of SMC SNR E0102]{ATCA Study of Small Magellanic Cloud Supernova Remnant 1E\,0102.2--7219}
\author[R. Z. E. Alsaberi et al. ]{Rami Z. E. Alsaberi,$^{1}$\thanks{E-mail: 19158264@student.westernsydney.edu.au}
M. D. Filipovi\'c,$^{1}$
S. Dai,$^{1,2}$
H. Sano,$^{3,4}$
R. Kothes,$^{5}$
J. L. Payne,$^{1}$
L. M. Bozzetto,$^{1}$
\newauthor
R. Brose,$^{6}$
C. Collischon,$^{7}$
E. J. Crawford,$^{1}$
F. Haberl,$^{8}$
T. Hill,$^{1}$
P. J. Kavanagh,$^{9}$
J. Knies,$^{7}$
D. Leahy,$^{10}$
\newauthor
P. J. Macgregor,$^{1,2}$
P. Maggi,$^{11}$
C. Maitra,$^{8}$
P. Manojlovi\'c,$^{1,2}$
S. Mart\'in,$^{12,13}$
C. Matthew,$^{1}$
N. O. Ralph,$^{1}$
\newauthor
G. Rowell,$^{14}$
A. J. Ruiter,$^{15}$
M. Sasaki,$^{7}$
I. R. Seitenzahl,$^{15}$
K. Tokuda,$^{16,17,18}$
N. F. H. Tothill,$^{1}$
\newauthor
D. Uro\v sevi\' c,$^{19,20}$
J. Th. van Loon,$^{21}$
V. Velovi\' c,$^{1}$
and F. P. A. Vogt$^{22}$
 \\
\\
Affiliations are listed at the end of the paper
}

\date{Accepted XXX. Received YYY; in original form ZZZ}

\pubyear{2021}

\begin{document}
\label{firstpage}
\pagerange{\pageref{firstpage}--\pageref{lastpage}}
\maketitle

\begin{abstract}
We present new and archival Australia Telescope Compact Array and Atacama Large Millimeter/submillimeter Array data of the Small Magellanic Cloud supernova remnant 1E\,0102.2--7219 at 2100, 5500, 9000, and 108000\,MHz; as well as \HI\ data provided by the Australian Square Kilometre Array Pathfinder. The remnant shows a ring-like morphology with a mean radius of %$\sim$
6.2\,pc. The 5500\,MHz image reveals a bridge-like structure, seen for the first time in a radio image. This structure is also visible in both optical and X-ray images. In the 9000\,MHz image we detect a central feature that has a flux density of 4.3\,mJy but rule out a pulsar wind nebula origin, due to the lack of significant polarisation towards the central feature with an upper limit of 4\,per\,cent. The mean fractional polarisation for 1E\,0102.2--7219 is $7\pm1$ and $12\pm2$\,per\,cent for 5500 and 9000\,MHz, respectively. The spectral index for the entire remnant is $-0.61\pm0.01$. We estimate the line-of-sight magnetic field strength in the direction of 1E\,0102.2--7219 of $\sim$44\,$\mu$G with an equipartition field of $65\pm5\,\mu$G. This latter model, uses the minimum energy of the sum of the magnetic field and cosmic ray electrons only. We detect an \HI\ cloud towards this remnant at the velocity range of $\sim$160--180\,km\,s$^{-1}$ and a cavity-like structure at the velocity of 163.7--167.6\,km\,s$^{-1}$. We do not detect CO emission towards 1E\,0102.2--7219.

%We present two different estimates of the magnetic field strength in the direction of 1E\,0102.2--7219. One estimate is line-of-sight, based on our polarisation rotation measure is $\sim$44\,$\mu$G. The other estimate, $65\pm5\,\mu$G, is based on an equipartition model calculated using intrinsic properties of the SNR as presented here. This latter model, uses the minimum energy of the sum of the magnetic field and cosmic ray electrons only.

% Using two ATCA observations seperated by 21 years we have measured the expansion rate of the small Magellanic cloud (SMC) super nova remnant (SNR) E0102.
\end{abstract}

\begin{keywords}
ISM: individual objects -- ISM: supernova remnants -- (galaxies:) Magellanic Clouds -- Radio continuum: ISM 
%{\bf UPDATE this and make it proper MNRAS style!}
\end{keywords}

%%%%%%%%%%%%%%%%%%%%%%%%%%%%%%%%%%%%%%%%%%%%%%%%%%

%%%%%%%%%%%%%%%%% BODY OF PAPER %%%%%%%%%%%%%%%%%%

\section{Introduction}

The Small Magellanic Cloud (SMC) is a gas-rich irregular dwarf galaxy orbiting the Milky Way (MW) with a current star formation rate of 0.021--0.05\,M$_{\odot}$~yr$^{-1}$ \citep{2018MNRAS.480.2743F}. As the second nearest star-forming galaxy after the Large Magellanic Cloud (LMC), its relatively small distance of $\sim$60\,kpc \citep{2005MNRAS.357..304H,2020ApJ...904...13G} and low Galactic foreground absorption ($N_{\rm HI}\sim6\times10^{20}$\,cm$^{-2}$) enables supernova remnants (SNRs) to be studied in great detail \citep{2008SerAJ.177...61C}. 
%entire population in the SMC to be studied in detail. various wavebands and at great spatial resolution. These qualities make the 
%SMC an ideal galaxy for observing celestial objects such as supernova remnants 
%\citep[SNRs;][]{2008SerAJ.177...61C}.
These remains of stellar explosions are characterised by strong non-thermal radio-continuum emission \citep[e.g.][]{2021map..book....2H,2023MNRAS.518.2574B} whose analysis provides a greater understanding of stellar evolution and the distribution of heavy elements \citep[e.g.][]{2021pma..book.....F}. %having a profound effect on the  morphology and evolution of the interstellar medium \citep[ISM;][]{2009SerAJ.179...55C}. %This evolution  is  highly dependent on  local environment as observed, for example, with the appearance of shell-like filaments from local interactions. 
%disturbed by local interaction and the inhomogeneous structure of the ISM.

%2021map..book....6M,2021MNRAS.500.2336Y

% \textcolor{red}{SNRs play a vital role in the physical evolution and chemical enrichment of the interstellar medium (ISM). Their precursor, SNe are believed to occur through two main scenarios. The first being core collapse events, which are the explosions of massive stars (M~$>$ 8--10 M$_{\odot}$), and release large quantities of heavy elements into the ISM. Alternatively, thermonuclear SNe (type\,Ia) progenitors are less massive and are believed to be the resulting detonation of carbon-oxygen white dwarfs that have reached the Chandrasekhar limit ($\sim$1.4 M$_{\odot}$). The study of type\,Ia SNe is not only important due to their use as `standard candles' in measuring cosmological distances, but also to ascertain the exact nature of the progenitor system.
% }

%E0102 is a well-studied SNR in optical, radio, and X-ray wavelengths. Located ~50 Kpc and 15 pc in projected distance north-east of the edge of the star forming region LHA 115-N76C \citep{henize1956catalogues}. It is one of the youngest SNR's known with a kinematic age of approximately 1000 yr.

1E\,0102.2--7219 (hereafter referred to as E0102) is a young (1738$\pm$175\,yr) core-collapse SMC SNR \citep{1988ApJ...327..156V,2021ApJ}, discovered via X-ray observations taken by the \textit{Einstein Observatory} \citep{1981ApJ...243..736S}. At X-ray wavelengths, E0102 presents a bright, filled ring-like structure with an outer edge that traces the forward-moving blast wave \citep{2005ApJ...632L.103S}. The bright X-ray ring shows strong emission lines of O, Ne and Mg, which are in the region between the reverse shock and the contact discontinuity with the ISM, revealing significant substructure \citep{2000ApJ...534L..47G,2006ApJ...642..260S}. E0102 is thought to originate from a core-collapse supernova scenario, given its `oxygen-rich' nature observed in the optical band \citep{1981ApJ...248L.105D}. \cite{2004ApJ...605..230F} calculated the  oxygen mass in the ejecta to be 6\,\(M_\odot\) using \cxo\ X-ray data, estimating a total progenitor mass of 32\,\(M_\odot\). 
Progenitor mass estimates from $\sim$40\,\(M_\odot\) using data from the \cxo\ archive \citep{2019ApJ...873...53A} to greater than 50\,\(M_\odot\) using Hubble Space Telescope (HST) data \citep{2006ApJ...641..919F} have been suggested.

\cite{1993ApJ...411..761A} published the first radio study of E0102, presenting two images using the Molonglo Observatory Synthesis Telescope (MOST) at 843\,MHz and the Australia Telescope Compact Array (ATCA) at 4790\,MHz. %Their ATCA image has a resolution of $2.75\times2.96$\,arcsec$^2$ and a root mean square (RMS) noise of $\sim$75\,$\mu$Jy\,beam$^{-1}$. 
%The ATCA image, having a resolution of $2.75\times2.96$\,arcsec$^2$ and a root mean square (RMS) noise of $\sim$75\,$\mu$Jy\,beam$^{-1}$, showed a $\sim$40\,arcsec diameter ring-like source.  Using their data, they estimated a radio  spectral index of $-0.7$. 
The ATCA image showed a $\sim$40\,arcsec diameter ring-like source (resolution = $2.75\times2.96$\,arcsec$^2$, root mean square (RMS) noise = $\sim$75\,$\mu$Jy\,beam$^{-1}$) and the authors estimated a radio spectral index of $-0.7$.

Here, we present new high-resolution and high-sensitivity radio-continuum observations of E0102 obtained from ATCA and the Atacama Large Millimeter/submillimeter Array (ALMA). 
%ATCA images, taken over the 2019-2021 period, are presented in Section~\ref{atca} \& Figs.~\ref{colored_2100}, \ref{colored1}, and \ref{colored2}. We combine these data with archival ATCA data and also present the first radio polarisation and \HI\ study of  E0102. 
%{\bf In Section~\ref{atca}, we present the ATCA images taken between 2019 and 2021, which are displayed in Figs.~\ref{colored_2100}, \ref{colored1}, and \ref{colored2}. Subsequently, these data are merged with archival ATCA data to conduct the first investigation into radio polarisation and \HI\ emission of E0102.}
%\HI\ data provided by Australian Square Kilometre Array Pathfinder (ASKAP) allows a better understanding of the environment surrounding E0102.
The general flow of our study of E0102 is as follows.  Section~\ref{obs} is a description of radio observations, data acquisition and initial analysis  from the  ATCA, Australian Square Kilometre Array Pathfinder (ASKAP), {\it Chandra}, HST, Multi-Unit Spectroscopic Explorer (MUSE), and ALMA telescopes. Next, Section~\ref{res} explores further analysis and results including radio morphology, polarisation, spectral index, rotation measure, magnetic field, and \HI\ / CO morphology. Section~\ref{com} is a discussion of the implications from our analysis, including the environment of E0102 and a comparison with other young SNRs. Final thoughts and conclusions are given in Section~\ref{con}.
%\textbf{(As I work through the paper, I may suggest this last paragraph need further modification.)}
%We present new high resolution and high sensitivity radio-continuum images of E0102 using combined data from our new ATCA observations on 2019, 2020, and 2021 with ATCA archival data. We also present the first radio polarisation and \HI\ study of this SNR. The structure of the paper is as follows: Section~\ref{obs} describes the observations and data reduction while we list our results and discussion Section~\ref{res}. In Section~\ref{com} we compare E0102 with similar Galactic and LMC SNRs. Finally, our conclusions are presented in Section~\ref{con}.

% The aim of this paper is study and measure the expansion rate of SNR E0102 shell observed with the Compact Array (ATCA) between 1992 and 2013 (21 years). Expansion is measured using an alternate machine vision inspired approach. Additionally, high-resolution polarisation images from 1997 ACTA observations will also be analysed.
  
%\citet{Others2013}, %\citep[e.g.][]{Author20
% NOTE: I may suggest more or less in the intro as I work through the paper.

\section{Observations} \label{obs}
\subsection{ATCA observations}\label{atca}

Our new (2019-2021) and archival\footnote{Australia Telescope Online Archive (ATOA), hosted by the Australia Telescope National Facility (ATNF): \url{https://atoa.atnf.csiro.au}} ATCA observations are listed in Table~\ref{tab:summary_obs}, including: observing date, project code, array configuation, number of channels, bandwidth, frequency, phase and flux calibrator used, and integrated time.
%--%We analysed ATCA observations
%; projects C2521, CX310, CX403, C3293, C3275, and C3296. These observations were taken on 2012~January~04$^\mathrm{th}$, 2015 January~04$^\mathrm{th}$, 2017~December~22$^\mathrm{th}$, 2019~Jun~20$^\mathrm{th}$, 2019~December~04$^\mathrm{th}$, 2020~Feb~21$^\mathrm{th}$, and 2021~Feb~26$^\mathrm{th}$, respectively. The observations 
%--%that are summarised in 
All  were carried out in `snap-shot' mode, with 1-hour integrations over a 12-hour period minimum, using the Compact Array Broadband Backend (CABB) (2048\,MHz bandwidth)  centred at wavelengths of 3/6\,cm \footnote{$\nu$~=~4500--6500 and 8000--10000\,MHz; centred at 5500 and 9000\,MHz, respectively} and 13\,cm ($\nu$~=~2100\,MHz). Total integration times were $\sim$575\,minutes and $\sim$1410\,minutes, respectively.  Complementary array configurations (Table~\ref{tab:summary_obs}) achieved good $uv$ coverage, including long  (6A, 6C, and 6D),  mid (1.5C), and  short (EW367) baselines, essential to imaging the full extent of E0102. The result of using such configurations is that our shortest baseline measured just 45.9\,m\footnote{see ATCA users guide, Appendix\,H: \url{https://www.narrabri.atnf.csiro.au/observing/users_guide/html/atug.html\#ATCA-Array-Configurations} for the baselines of each array configuration} (array EW367) and the longest reaching 5938.8\,m (array 6A). In contrast, \cite{1993ApJ...411..761A} only utilised five of the six antennas, with a minimum baseline of 76.5\,m and the longest at 2525.5\,m -- which significantly affected the quality of the image given the limited $uv$ coverage.%\textbf{In fact, our latest observations feature an extensive range of baselines, with the longest reaching 5938.8-m\footnote{see ATCA users guide, Appendix\,H: \url{https://www.narrabri.atnf.csiro.au/observing/users_guide/html/atug.html\#ATCA-Array-Configurations} for the baselines of each array configuration} (array 6A) and the shortest measuring 45.9-m (array EW367). In contrast, \cite{1993ApJ...411..761A} utilised five out of six antennas, which significantly affected the quality of the image given their poor $uv$ coverage. They employed OLD\,3 and 6\,km arrays (without the 6$^\mathrm{th}$ antenna) where only one baseline is below 100-m (at 76.5-m) and the longest at 2525.5-m.} %The total amount of time on source was 194\,minutes.  }

%\textcolor{red}{The result of using such configurations is that our shortest baseline measured just 45.9\,m\footnote{see ATCA users guide, Appendix\,H: \url{https://www.narrabri.atnf.csiro.au/observing/users_guide/html/atug.html\#ATCA-Array-Configurations} for the baselines of each array configuration} (array EW367) and the longest reaching 5938.8\,m (array 6A). In contrast, \cite{1993ApJ...411..761A} only utilised five of the six antennas, with a minimum baseline of 76.5\,m and the longest at 2525.5\,m -- which significantly affected the quality of the image given the limited $uv$ coverage.}

%Source PKS\,B1934--638 was used as the primary (flux density) calibrator for all observations, except on 2017~December~22$^\mathrm{th}$ the primary calibrator was PKS\,B0252--721. While, source PKS\,B0230--790 was used as the secondary (phase) calibrator for all observations, except on 2019~December~04$^\mathrm{th}$ and 2021~Feb~26$^\mathrm{th}$, the secondary calibrator was j0047--790 and PKS\,B0530--727, respectively. %However, pre-CABB (C33) (described in \citet{1993ApJ...411..761A}) observations were carried on in full imaging mode.

We used \textsc{miriad}\footnote{\url{http://www.atnf.csiro.au/computing/software/miriad/}} \citep{1995ASPC...77..433S}, \textsc{karma}\footnote{\url{http://www.atnf.csiro.au/computing/software/karma/}} \citep{1995ASPC...77..144G}, and \textsc{DS9}\footnote{\url{https://sites.google.com/cfa.harvard.edu/saoimageds9}} \citep{2003ASPC..295..489J} software packages for reduction and analysis. All observations were calibrated using the phase and flux calibrators as listed in Table~\ref{tab:summary_obs} with one iteration of phase-only self-calibration using the \textsc {selfcal} task. To best image the diffuse emission emanating from the SNR, we experimented with a range of various values for weighting and tapering. We found Briggs weighting robust parameter of --2 (uniform weighting) the most optimal choice. We add a Gaussian taper to further enhance the diffuse emission (1\,arcsec at 5500\,MHz and 1.5\,arcsec at 9000\,MHz) which resulted in a resolution of $2\times2$\,arcsec$^{2}$ for these images. %Also, an additional 1\,arcsec and 1.5\,arcsec Gaussian taper is included to further enhance the diffuse emission for the 5500\,MHz and 9000\,MHz images, respectively. This allowed us to achieve a resolution of $2\times2$\,arcsec$^{2}$ for the 5500 and 9000\,MHz images.
The \textsc {mfclean} and \textsc {restor} algorithms were used to deconvolve the images, with primary beam correction applied using the \textsc{linmos} task. To increase the S/N ratio for weak polarisation emission, we follow the same process with stokes {\it Q} and {\it U} parameters but with a beam size of $5\times5$\,arcsec$^2$ (see Section~\ref{pol} for more details). %Pre-CABB projects C33 has a significantly lower bandwidth, which were deconvolved using only the Steer CLEAN "mossdi" task. CABB projects with a wider bandwidth was deconvolved using the "mfclean" task.

%\textcolor{red}{We add a Gaussian taper to further enhance the diffuse emission (1\,arcsec at 5500\,MHz and 1.5\,arcsec at 9000\,MHz) which resulted in a resolution of $2\times2$\,arcsec$^{2}$ for these images.}

% Polarisation analysis was completed by additionally synthesising and deconvolving stokes Q and U. We produced polarisation intensity and fractional polarisation maps by using the \textsc{impol} task, the polarised intensity and position angles between stokes Q and U are computed. %We used \textsc{cgdisp} to display the results as polarisation vectors superimposed on the original image. %We emphasise that observations from ATCA project C33 did not have polarisation capability.  

\begin{table*}
	\centering
	\caption{Summary of SNR E0102 ATCA observations used in this study.}
	\label{tab:summary_obs}
	\begin{tabular}{@{}clccllllr@{}}
		\hline
    Observing   & Project& Array    & No.       & Bandwidth &  Frequency $\nu$  & Phase         & Flux              & Integrated time   \\
    Date        & Code   & Config.  & Channels  &  (GHz)    &  (MHz)        & Calibrator        & Calibrator        & (minutes) \\
		\hline
    %   20$^{th}$ Mar~1992   & C33   & 6A                & 33       &  0.128    &   4790,8640   & 0252.712     & 1934-638\\ 
    %   31$^{st}$ Mar~1992   & C33   & 6C                & 33       &  0.128    &   4790,8640   & 0252.712     & 1934-638\\
    %   05$^{th}$ Jun~1992   & C33   & 1.5D              & 33       &  0.128    &   4790,8640   & 0252.712     & 1934-638\\
    %   18$^{th}$ Jun~1992   & C33   & 1.5B              & 33       &  0.128    &   4790,8640   & 0252.712     & 1934-638\\
    04~Jan~2012 & C2521 & 6A        & 2049      &  2.048    &   2100        & PKS\,B0230--790   & PKS\,B1934--638   & 1410.1\\
    04~Jan~2015 & CX310 & 6A        & 2049      &  2.048    &   5500,9000   & PKS\,B0230--790   & PKS\,B1934--638   & 37.8 \\
    22~Dec~2017 & CX403 & 6C        & 2049      &  2.048    &   5500,9000   & PKS\,B0230--790   & PKS\,B0252--712   & 193.2 \\
    20~Jun~2019 & C3293 & 6A        & 2049      &  2.048    &   5500,9000   & PKS\,B0230--790   & PKS\,B1934--638   & 119.4 \\
    04~Dec~2019 & C3275 & 1.5C      & 2049      &  2.048    &   5500,9000   & J0047--7530       & PKS\,B1934--638   & 46.2 \\
    21~Feb~2020 & C3296 & EW367     & 2049      &  2.048    &   5500,9000   & PKS\,B0230--790   & PKS\,B1934--638   & 75.6  \\
    26~Feb~2021 & C3403 & 6D        & 2040      &  2.048    &   5500,9000   & PKS\,B0530--727   & PKS\,B1934--638   & 102.6\\
		\hline
	\end{tabular}
\end{table*}

\subsection{\texorpdfstring{\HI}{H} observations}
 The \HI\ data used in this study are from \cite{2022PASA...39....5P}. These \HI\ data (1420.4\,MHz) were obtained using the ASKAP with an angular resolution of $30\times30$\,arcsec$^2$, corresponding to a spatial resolution of $\sim$9\,pc at the distance of the SMC. The typical noise fluctuations of the \HI\ data cube are $\sim$1.1\,K at a velocity resolution of 0.98\,km\,s$^{-1}$. 
%\footnote{\url{https://doi.org/10.25919/www0-4p48}}
\subsection{Chandra observations}\label{chandra}
We analysed archival X-ray data obtained using \cxo\ with the Advanced CCD Imaging Spectrometer. We combined 142 individual observations from August 1999 (ObsID: 138) to February 2021 (ObsID: 24579) using Chandra Interactive Analysis of Observations \citep[CIAO;][]{2006SPIE.6270E..1VF} software version 4.12 with CALDB 4.9.1 \citep{2007ChNew..14...33G}. The data were reproduced using the \textsc{chandra\_repro} procedure. Using the `merge\_obs' procedure in the energy band of 0.5--7.0\,keV, we produced an exposure-corrected, energy-filtered image. The total effective exposure time is $\sim$1570\,ks.

\subsection{HST observations}\label{hst}
The \OIII\ filter HST image of  E0102 was downloaded from the Mikulski Archive for Space Telescopes\footnote{\url{https://mast.stsci.edu/portal/Mashup/Clients/Mast/Portal.html}}. Information about these HST observations and their data reduction are detailed in \cite{2000ApJ...537..667B}.

\subsection{MUSE observations}
The [Fe\,\textsc{xiv}] image of E0102 was created as a slice through the IFU datacube from observations on 7$^\mathrm{th}$~October 2016 using MUSE under ESO program 297.D-5058 (PI: F. P. A. Vogt). All details regarding these MUSE observations and their data reduction are found in \cite{2017A&A...602L...4V}.

\subsection{ALMA observations}
We utilised archival datasets from both continuum and CO line emissions obtained using  ALMA. The 3\,mm continuum observations centred at 108058\,MHz were carried out using the 12-m array as a cycle 6 project (PI: F. Vogt, \#2018.1.01031.S). 
% Correlator was setup with four spectral windows 1875\,MHz centered at 101125, 103000, 113125, and 115000\,MHz.
The correlator was set with four spectral windows 1875\,MHz wide, centred at 101125, 103000, 113125, and 115000\,MHz. Observations consisted of two executions with 48 and 45 antennas in a compact array configuration with baselines ranging between 15.1 and 500.2\,m which correspond to maximum recoverable scales of $\sim$16\,arcsec \citep[following Eq. 7.7 in the ALMA Cycle 8 Technical Handbook\footnote{\url{https://almascience.nrao.edu/documents-and-tools/cycle8/alma-technical-handbook}}][]{2021athb.rept.....R}. We used the products delivered by the pipeline  processed with the Common Astronomy Software Application \citep[CASA;][]{2007ASPC..376..127M} package version 5.4.0-70 and Pipeline version 42254 (Pipeline-CASA54-P1-B). Imaging incorporated a Briggs robust value of 1.0. The ALMA radio continuum image % was obtained by over 
covers a total bandwidth of 7290\,MHz and achieved an RMS of 31\,$\mu$m. The beam size is $2.46\times1.93$\,arcsec$^2$ with a position angle of 0\fdg73 corresponding to a spatial resolution of $\sim$0.6\,pc at the distance of the SMC. Additional observations with a more extended configuration were available in the archive but since no continuum was detected at higher resolution, they were not included. Spectral line emission  $^{12}$CO (J~=~1--0), covered by one of the spectral windows in this dataset, is not detected.
% in this dataset.

%\footnote{\url{https://almascience.nrao.edu/documents-and-tools/cycle8/alma-technical-handbook}}
Since spatial filtering may be significant \citep[see section 3.3 in the ALMA Cycle 8 Technical Handbook of][]{2021athb.rept.....R}, the line emission $^{12}$CO (J~=~2--1) was obtained with the Atacama Compact Array (ACA, a.k.a. Morita Array) comprising the 7-m ALMA antennas and 12\,m total power array, carried out as a wide field survey in Cycle 5/6 (PI: C. Agliozzo, \#2017.A.00054.S). The baselines ranging from 9 to 48\,m result in maximum recoverable scales of 27\,arcsec. Additionally, this survey allows us to explore a wider region, comparable to the \HI\ dataset (see Section~\ref{HI}). The whole dataset, beyond the region used in this study, has been published by \cite{2021ApJ...922..171T}. Details on data reduction, imaging and combination between interferometric data and single-dish observations can be found in \cite{2021ApJ...922..171T}. The final beam size is $6.87\times 6.56$\,arcsec$^2$ with a position angle of 39\fdg09, corresponding to a spatial resolution of $\sim$2\,pc at the distance of the SMC. The typical noise fluctuation is $\sim$66\,mK at the velocity resolution of 0.5\,km\,s$^{-1}$.

\begin{figure*}
     \centering
     \includegraphics[scale=0.35, trim=0 0 0 0,clip]{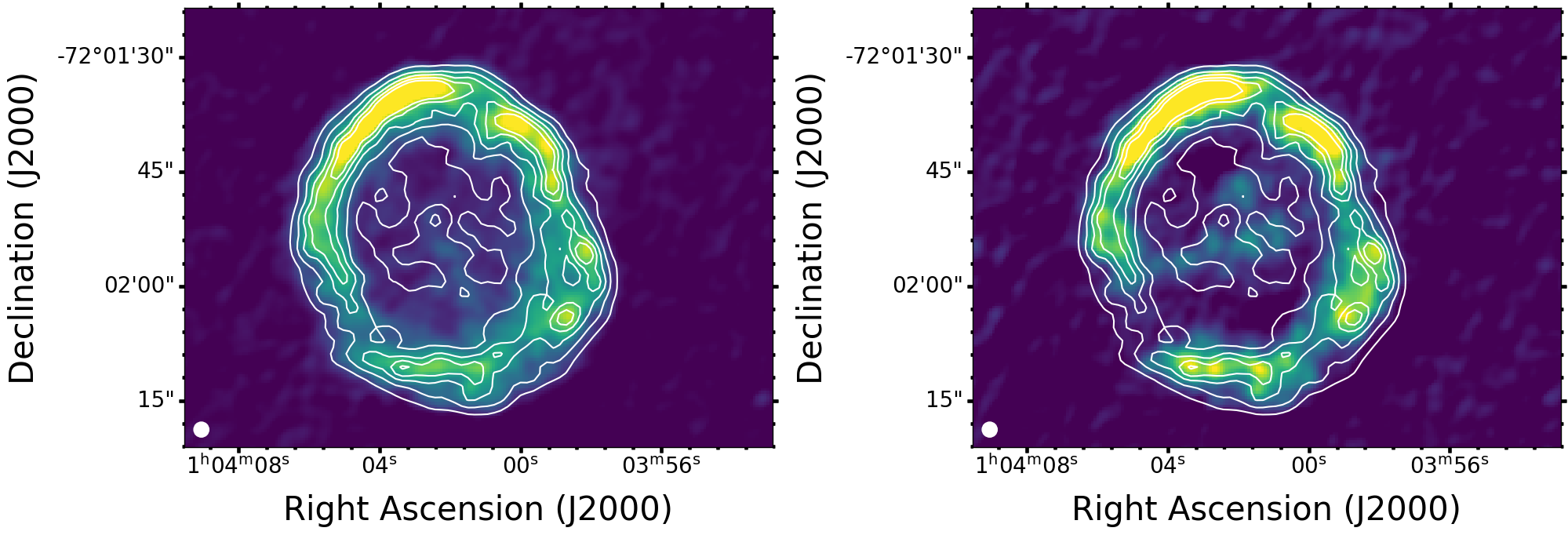}
%     %\includegraphics[scale=0.44, trim=0 0 0 0,angle=270,clip]{colerd_5500_new.eps}
%     %\includegraphics[scale=0.44, trim=0 40 0 0 0,angle=270,clip]{colored_9000_new.eps}
    \caption{Images of SNR E0102 with 2100\,MHz contours overlaid from 500 to 2500 $\mu$Jy\,beam$^{-1}$ at increments of 500\,$\mu$Jy\,beam$^{-1}$. \emph{Left:} ATCA intensity image at 5500\,MHz. \emph{Right:} ATCA image at 9000\,MHz. The white circles on the lower left corner in both images represent a synthesised beam of $2\times2$\,arcsec$^2$. } % The contours are from ATCA image (200, 400, 600, 800, and 1000\,$\mu$J\,beam$^{-1}$). The white circle at the lower left corner of ATCA image represents a synthesised beam of $2\times2$\,arcsec.}
    \label{colored_2100}
\end{figure*}
\begin{figure*}
     \centering
     \includegraphics[scale=0.33, trim=0 0 0 0,clip]{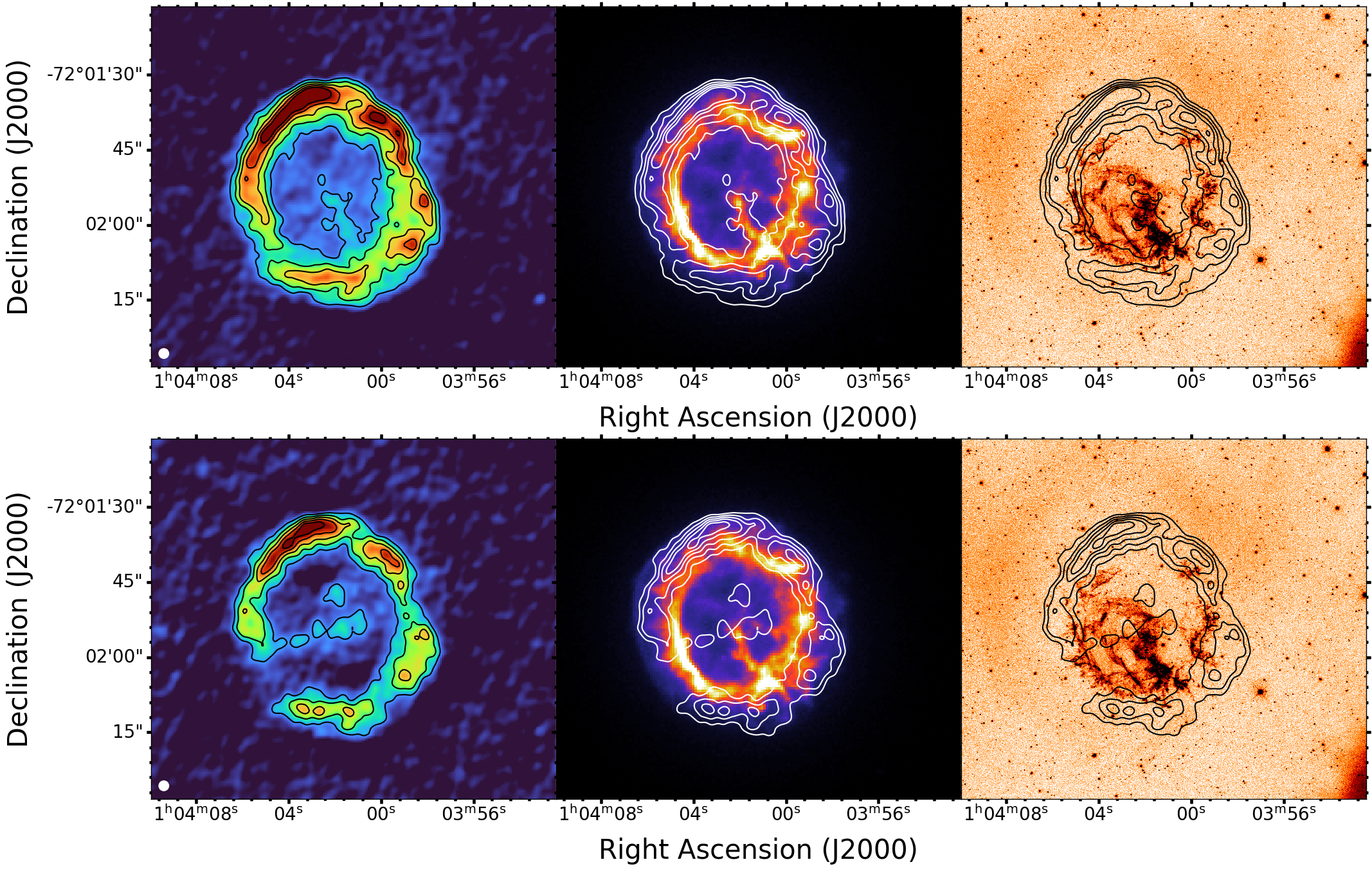}
%     %\includegraphics[scale=0.44, trim=0 0 0 0,angle=270,clip]{colerd_5500_new.eps}
%     %\includegraphics[scale=0.44, trim=0 40 0 0 0,angle=270,clip]{colored_9000_new.eps}
    \caption{Images of SNR E0102 with 5500\,MHz contours overlaid (top) and 9000\,MHz contours overlaid (bottom). The contour levels are from 200 to 1000 $\mu$Jy\,beam$^{-1}$ at increments of 200\,$\mu$Jy\,beam$^{-1}$. \emph{Left:} ATCA intensity image at 5500\,MHz (top) and 9000\,MHz (bottom). The white circles on the lower left corner represent a synthesised beam of $2\times2$\,arcsec$^2$. \emph{Middle:} Broadband \cxo\ X-ray images. \emph{Right:} HST \OIII\ filter images \citep{2000ApJ...537..667B}.}%The contours are from ATCA image (200, 400, 600, 800, and 1000 $\mu$J\,beam$^{-1}$). The  synthesised beam of $2\times2$\,arcsec$^2$ is show as a white circle in the lower left corner of the ATCA image.}% represents a synthesised beam of $2\times2$\,arcsec. }
    \label{colored4}

\end{figure*}

%\section{Results and Discussion} \label{res}
\section{Analysis} \label{res}

\subsection{Morphology}

\label{mor}

Our new ATCA images of E0102 clearly show a ring-like structure with bright emissions in the north-west and north-east regions (Fig.~\ref{colored_2100}). A comparison between the high-resolution radio continuum (ATCA) images of SNR E0102 at 5500 and 9000\,MHz and other wavelength images is presented in Fig. \ref{colored4}. The RMS noise for 5500\,MHz and 9000\,MHz images are $\sim$30 and 48\,$\mu$Jy\,beam$^{-1}$, respectively. In this Section, we describe how we calculate the centre and the radius of E0102, compare our new images with X-ray and optical images, and list new features seen for the first time in radio images.

\subsubsection{Centre and radius calculation} \label{centre}
We used the Minkowski tensor analysis tool BANANA\footnote{\url{https://github.com/ccollischon/banana}} \citep{2021A&A...653A..16C} to determine the centre of expansion. This tool searches for  filaments and calculates normal lines and line density maps. 
For a perfect shell, all lines should meet at the centre position inside the SNR, where the expansion must have started \citep[see][for more details]{2021A&A...653A..16C}. Since real sources deviate from this ideal picture, we circumvented this problem by smoothing the line density map using a circle with a diameter of 40 pixels. We then took the centre position of the pixel where the line density was the highest. Using the same method, we performed the tensor analysis separately for the 5500\,MHz, 9000\,MHz, and \textit{Chandra} broad-band images. From this, we calculated the mean centre position and the $1\sigma$ uncertainty of the mean.
The resulting calculated centre is RA~(J2000)~=~01$^{h}$04$^{m}$02.23$^{s}$, Dec~(J2000)~=~$-$72$^\circ$01$\arcmin$53\farcs06 with an uncertainty of $\sim$0.91\,arcsec. This is $\sim$1.5\,arcsec ($\sim$0.4\,pc at the distance of 60\,kpc) north-west from the previous estimation of RA~(J2000)~=~01$^{h}$04$^{m}$02.48$^{s}$, Dec~(J2000)~=~$-$72$^\circ$01$\arcmin$53\farcs92, having a $1\sigma$ uncertainty of 1.77\,arcsec \citep{2021ApJ}. The latter was derived from HST observations of the proper motion of optical ejecta. Therefore, our calculated position agrees well with previous estimates, within uncertainties.
%(see Fig.~\ref{fig:centres})
%Where all the lines meet inside the SNR (the line density is highest), that is where the expansion must have started \citep[see][for more details]{2021A&A...653A..16C}. The calculated centre for the radio continuum emission at 5500\,MHz is RA~(J2000)~=~01$^{h}$04$^{m}$02.35$^{s}$, Dec~(J2000)~=~$-$72$^\circ$01$\arcmin$52\farcs11. This is $\sim$2\,arcsec ($\sim$0.6\,pc at the distance of 60\,kpc) north-west from the previous estimation of RA~(J2000)~=~01$^{h}$04$^{m}$02.48$^{s}$, Dec~(J2000)~=~$-$72$^\circ$01$\arcmin$53\farcs92 \textbf{with $1\sigma$ uncertainty of 1.77\,arcsec} \citep{2021ApJ}, which is \textbf{derived from HST observations of the proper motion of the optical ejecta} (Fig.~\ref{fig:centres}). While, the calculated centre for the broadband \textit{Chandra} X-ray image is RA~(J2000)~=~01$^{h}$04$^{m}$02.40$^{s}$, Dec~(J2000)~=~$-$72$^\circ$01$\arcmin$53\farcs42 which is located $\sim$0.6\,arcsec ($\sim$0.17\,pc) north-west from the \cite{2021ApJ} position but still within their error circle (see Fig.~\ref{fig:centres}).

\begin{figure*}
\centering
\includegraphics[scale=1.3,trim=90 30 100 30,clip]{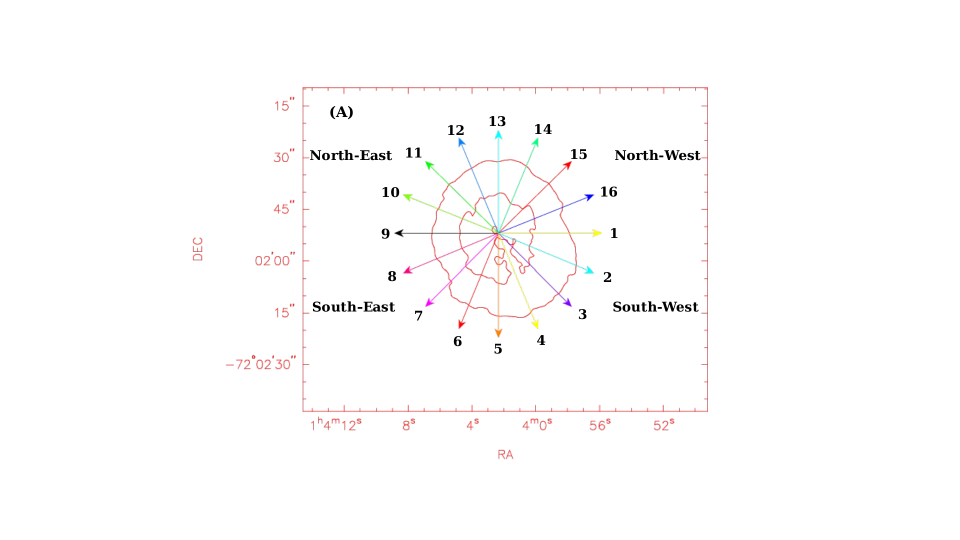}\\
\includegraphics[scale=0.280,trim=10 0 100 50,clip]{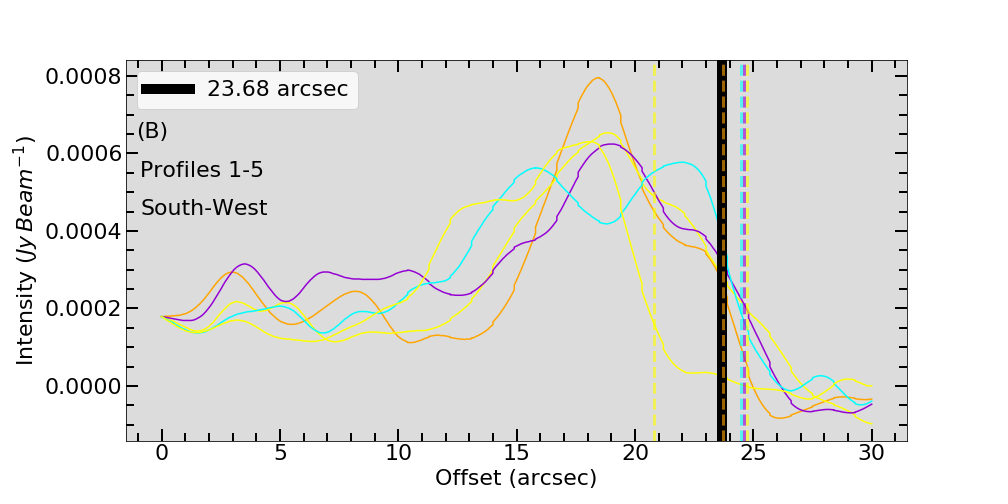}
\includegraphics[scale=0.280,trim=10 0 100 60,clip]{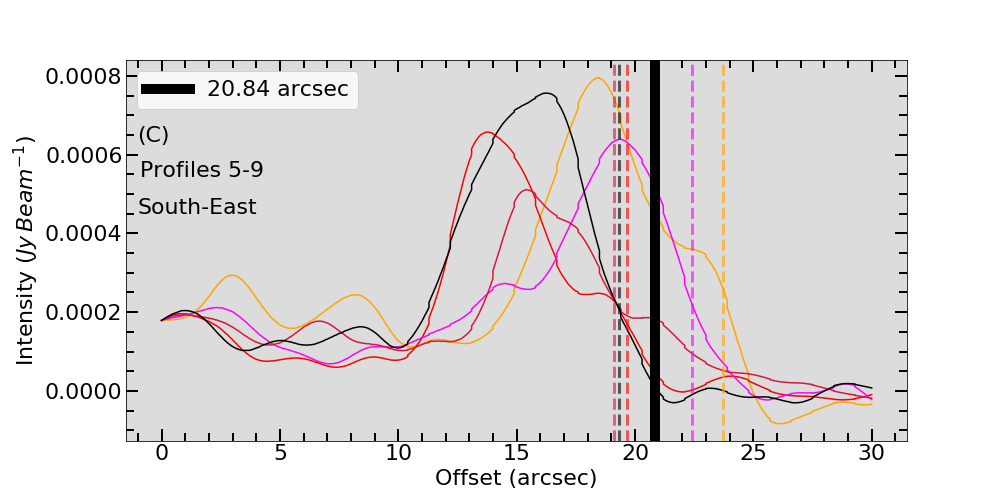}\\
\includegraphics[scale=0.280,trim=10 0 100 60,clip]{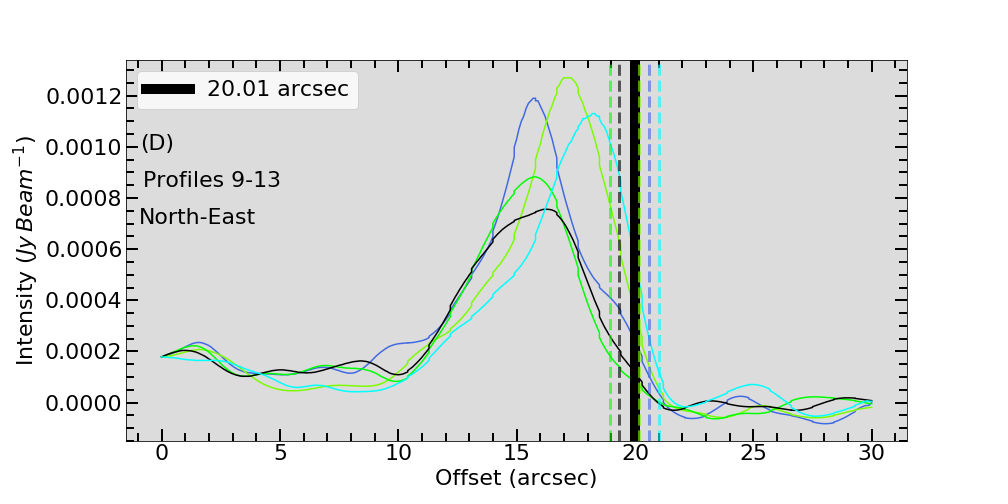}
\includegraphics[scale=0.280,trim=10 0 100 60,clip]{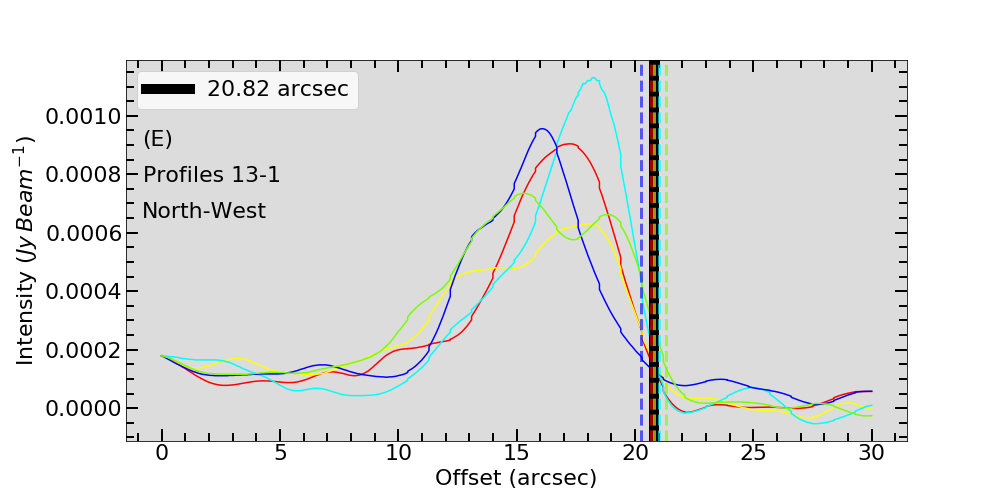}\\
\caption{Estimate of the radio continuum radius of E0102 at 5500\,MHz. A: Radial profiles around the remnant from the centre (see Section~\ref{mor}) overlaid on the $\bm{3\sigma}$ ATCA image contour (200\,$\mu$Jy\,beam$^{-1}$) at 5500\,MHz. The central position of the remnant is RA~(J2000)~=~01$^{h}$04$^{m}$02.35$^{s}$, Dec~(J2000)~=~$-$72$^\circ$01$\arcmin$52\farcs11. B, C, D, and E: Radial profiles. The dashed vertical lines represent profile cutoffs and the thick black vertical lines represent the average of the cutoffs (in arcsec) for different parts of the shell; south-west, south-east, and north-east, and north-west, respectively. }
%(S-W), north-west (N-W) \& north-east (N-E), and south-east (S-E). Each colour in this graph corresponds to the coloured vectors in the upper image.}
\label{fig:slices}
\end{figure*}

To estimate the radio continuum radius of E0102, we used the \textsc{miriad} task \textsc{cgslice} to plot 16 equispaced radial profiles in 22.5$^\circ$ segments around the remnant at 5500\,MHz. Each profile is 30\,arcsec in length (Fig.~\ref{fig:slices}). We divide the remnant into four equal regions: south-west (profiles 1--5), south-east (profiles 5--9), north-east (profiles 9--13), and north-west (profiles 13--1) (Fig.~\ref{fig:slices}, A). We identified the cutoff as the point where each profile intersects the outer contour line ($3\sigma$ ATCA image contour or 200\,$\mu$Jy\,beam$^{-1}$ at 5500\,MHz). These cutoffs are represented as dashed vertical lines (see Fig.~\ref{fig:slices}, B, C, D, and E). The thick black vertical lines represent the average of the cutoffs in each region.
%The vertical black lines (see Fig.~\ref{fig:slices}, B, C, and D) represent the average cutoff values of the slices for each region. 
The resulting radii vary from $23.7\pm01.7$\,arcsec ($6.9\pm0.5$\,pc) towards the south-west (Fig.~\ref{fig:slices}, B) to $20.9\pm02.1$\,arcsec towards the south-east ($6.1\pm0.6$\,pc; Fig.~\ref{fig:slices}, C) and $20.0\pm0.9$\,arcsec ($5.8\pm0.3$\,pc) towards the north-east (Fig.~\ref{fig:slices}, D) to $20.9\pm0.4$ ($6.1\pm0.1$\,pc) towards the north-west (Fig.~\ref{fig:slices}, E) with an average of $21.4\pm01.3$\,arcsec ($6.2\pm0.4$\,pc) for the entire SNR. 

\begin{figure*}
\centering
\includegraphics[scale=1.7,trim=4 0 0 0,clip]{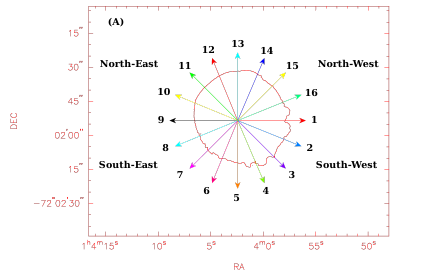}\\
\includegraphics[scale=0.275,trim=0 0 100 60,clip]{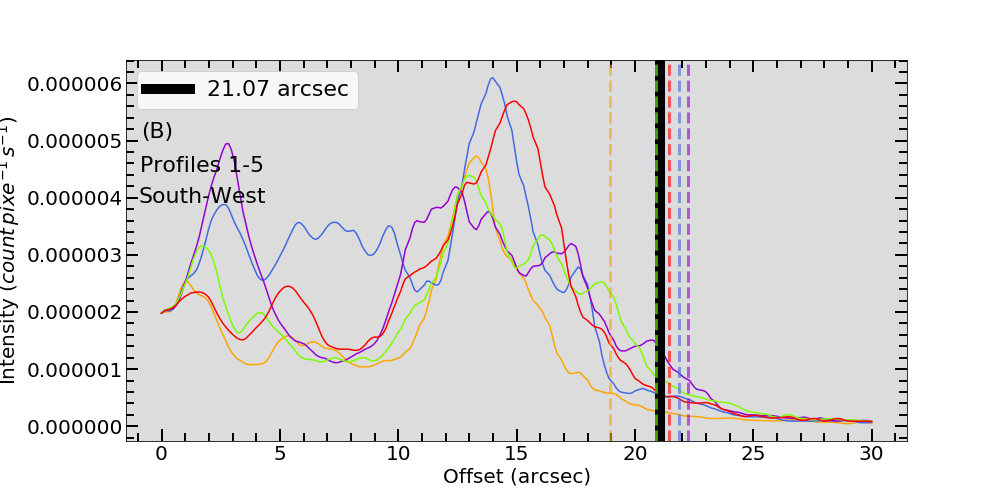}
\includegraphics[scale=0.275,trim=0 0 100 60,clip]{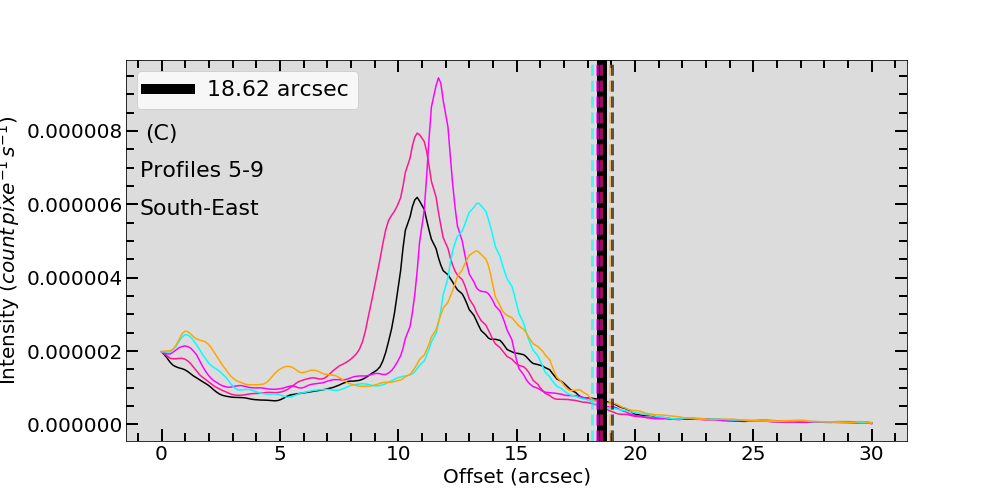}\\
\includegraphics[scale=0.275,trim=0 0 100 60,clip]{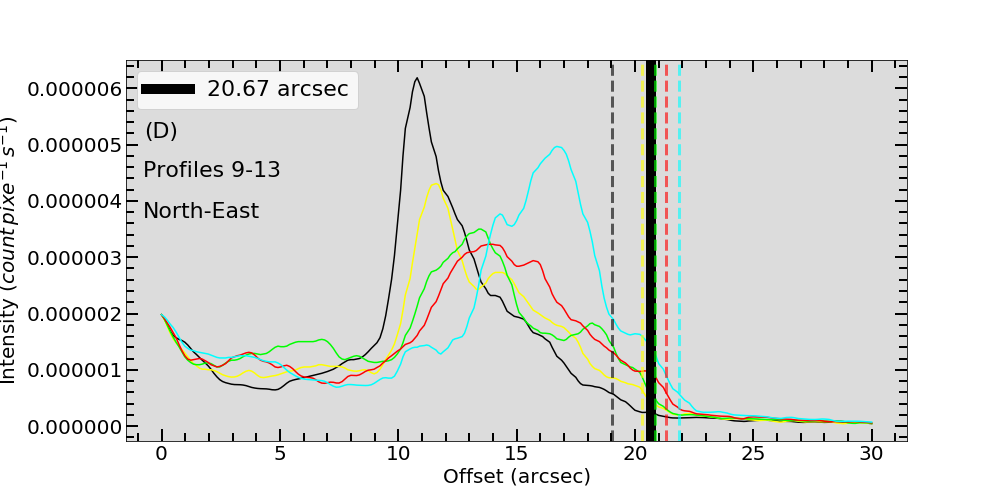}
\includegraphics[scale=0.275,trim=0 0 100 60,clip]{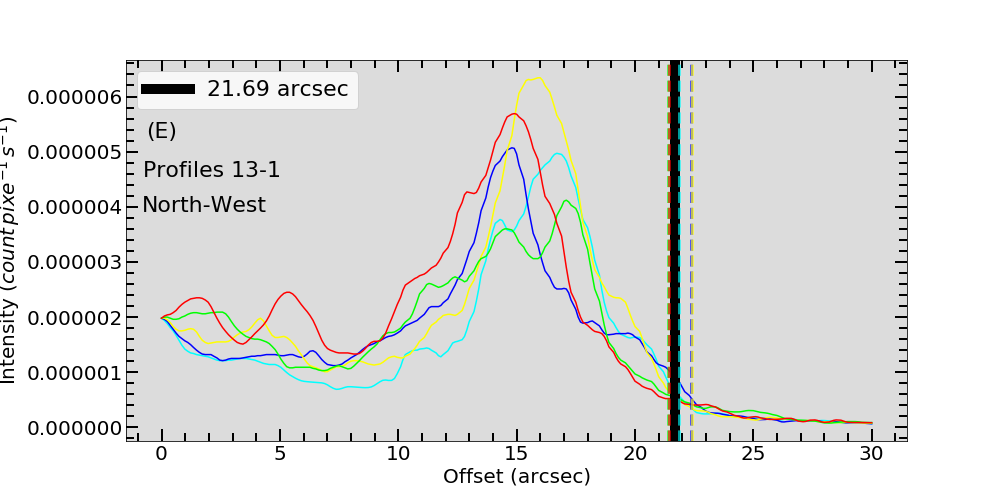}\\
\caption{Estimate of X-ray radius of E0102. A: Radial profiles overlaid on $3\sigma$ \cxo\ broadband X-ray image contour ($6\times10^{-7}$\,counts~pixel$^{-1}$s$^{-1}$). The central position of the remnant is RA~(J2000)~=~01$^{h}$04$^{m}$02.40$^{s}$, Dec~(J2000)~=~$-$72$^\circ$01$\arcmin$53\farcs42. B, C, D, and E: Radial profiles. The dashed vertical lines represent profile cutoffs and the thick black vertical lines represent the average of the cutoffs (in arcsec) for different parts of the shell; south-west, south-east, north-east, and north-west, respectively. }
%(S-W), north-west (N-W) \& north-east (N-E), and south-east (S-E). Each colour in this graph corresponds to the coloured vectors in the upper image.}
\label{slices.X-ray}
\end{figure*}

Using the same methodology we estimate the X-ray radii based on the \cxo\ image (Fig.~\ref{slices.X-ray}). 
The radii vary from $21.1\pm01.3$\,arcsec ($6.1\pm0.4$\,pc) towards the south-west (Fig.~\ref{slices.X-ray}, B) to $18.6\pm0.4$\,arcsec ($5.4\pm0.1$\,pc) towards the south-east (Fig.~\ref{slices.X-ray}, C) and $20.7\pm01.1$\,arcsec ($6.0\pm0.3$\,pc) towards the north-east (Fig.~\ref{slices.X-ray}, D) to $21.7\pm0.4$\,arcsec ($6.3\pm0.1$\,pc) towards the north-west (Fig.~\ref{slices.X-ray}, E) with an average of $20.5\pm0.8$\,arcsec ($5.1\pm0.2$\,pc).

 %The centre of the remnant is as per \citet{2021ApJ...912...33B}
%%%%%%%%%%%%%%%%%%%%%%%%%%%%%%%%%%%%%%%%%%%%%%%%%%%%%%%%%%%%%%%%%
%{\bf you should also do the same radial plot analysis for the X-ray and optical image --- and than, compare three plots. }
\subsubsection{Radio vs. X-ray vs. optical}
A comparison of our new ATCA data (Section~\ref{atca}) with the \cxo\ X-ray image (Section~\ref{chandra}) and optical HST image (Section~\ref{hst}) reveals that the
radio/X-ray maximum does not correlate well, while there is no apparent correlation with optical emission either.
%We compare our new radio images with \cxo\ X-ray image and optical (HST) image. Interestingly the X-ray and radio images are anti correlated, the maximum emissions in the radio image coincide with the minimum emissions in the X-ray image and vise versa, very little could be correlated with optical emission. 
In the similar age Galactic SNR Vela\,Jr (G266.2--1.2), \citet{2005AdSpR..35.1047S} and \citet{2018ApJ...866...76M} observed the opposite, where radio emission is trailing behind the X-ray emission. 

\begin{figure*}
     \centering
     \includegraphics[width=0.49\textwidth, trim=0 0 0 0,clip]{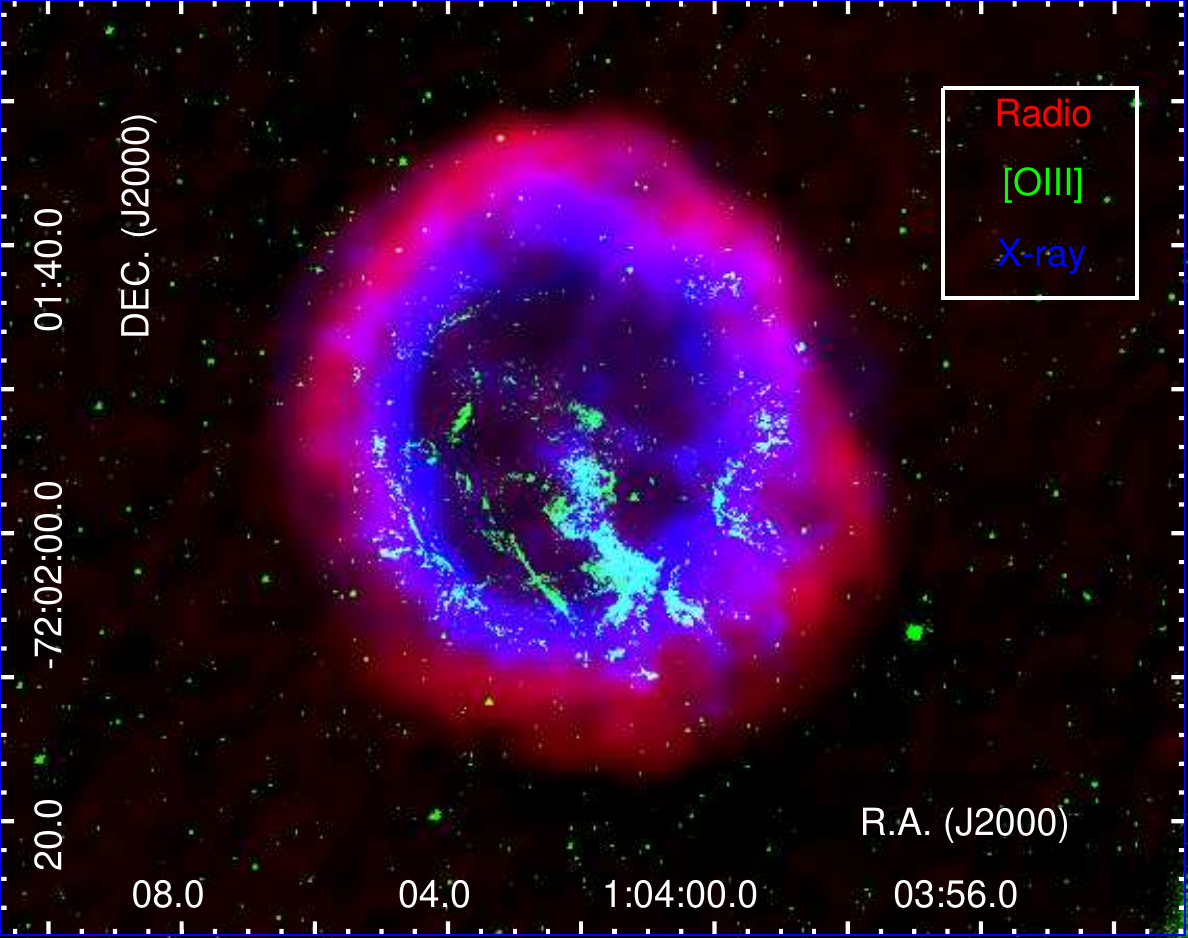}
      \includegraphics[width=0.49\textwidth, trim=0 0 0 0,clip]{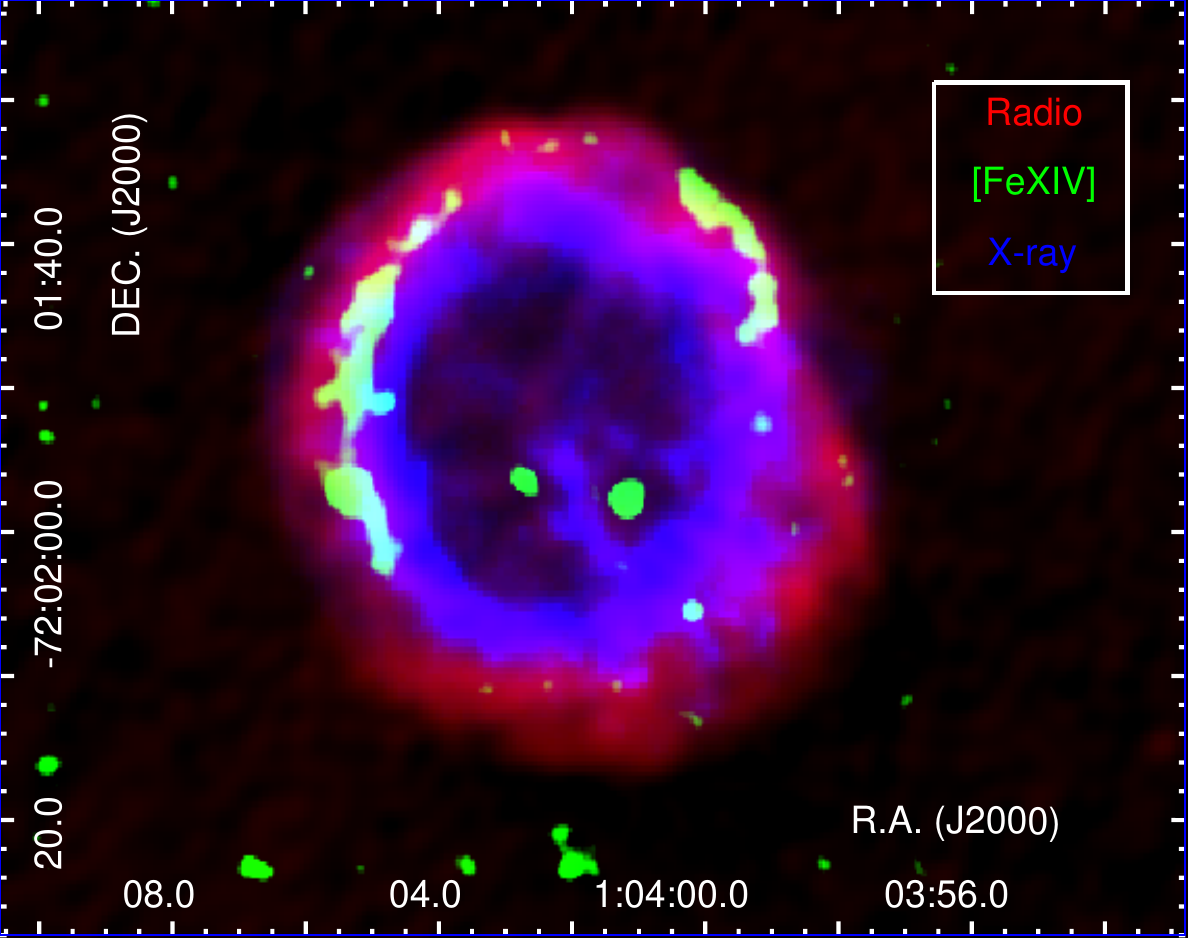}
     \caption{Red-Green-Blue (RGB) image of the E0102 SNR, with the key printed in the upper right-hand corner. \emph{Left:} Our 5500\,MHz ATCA data are in red, the HST image (\OIII\ filter) in green \citep{2000ApJ...537..667B}, and the \cxo\ broadband image is in blue. \emph{Right:} 5500\,MHz ATCA data are in red, the smoothed MUSE image ([Fe\,\textsc{xiv}]) in green \citep{2017A&A...602L...4V}, and the \cxo\ broadband image is in blue.}
      \label{colored3}
 \end{figure*}

Fig.~\ref{colored3} (left) %is a red-green-blue (RGB) image of E0102, with R being our new ATCA data at 5500\,MHz, G being the HST optical image (\OIII\ filter) \citep{2000ApJ...537..667B}, and B being the broadband \cxo\ X-ray image. This figure
shows radio emission extends beyond X-ray emission, as it most likely traces the forward shock while X-ray emission could be a tracer of reverse shocked gas \citep{2000ApJ...534L..47G}. Further, some of the radio profiles (e.g. 2, 6, 8) show secondary peaks offset towards the centre of the remnant. These second peaks, however, have a larger radius than the X-ray emission, indicating a second radio ring in parts of the remnant that can be associated with the amplified magnetic field that is expected at the contact-discontinuity (CD) between the shocked ejecta and shocked ISM \citep{1996ApJ...465..800J, 2014ApJ...785..130Z}.

Interestingly, we note that the [Fe\,\textsc{xiv}] emission is sandwiched between radio and X-ray emission (see Fig.~\ref{colored3}, right). We suggest we see two phases of clumpy medium: the forward shock (a fast blast wave of a few 1000\,km\,s$^{-1}$) and  denser parts of circumstellar medium (CSM)/ISM  (where the blast wave is driving slower cloud shocks). The [Fe\,\textsc{xiv}] emission is coming from shocks with velocities around 340\,km\,s$^{-1}$ \citep{2017A&A...602L...4V}. It takes some finite time for the shocked material in the denser clumps to be ionised behind the cloud shock giving us the iron lines we see; during that time the blast wave has already raced ahead, placing spatial separation between the two emitting regions. The two bright sources near the centre of the remnant and all point sources in the field are stars (Fig.~\ref{colored3}, right).

\subsubsection{Bridge like structure}
We note a striking similarity of a `bridge-like structure' feature between our 5500\,MHz (ATCA) and the \cxo\ image, %We However, we note a striking similarity between our 5500\,MHz and \cxo\ image of a feature `bridge like structure` 
which projects from the centre of the remnant and connects to the south-west side of the E0102 ring. However, it is comparatively less prominent in HST image (see Fig.~\ref{colored4}, upper panel).  
%Even more interestingly, we cannot detect this feature in addition to the south-east limb in our 9000\,MHz image.
This feature is not detected at 9000\,MHz and is likely the result of interactions with an interstellar cloud seen in \textit{Spitzer} and \textit{Herschel} data and similar to that seen in LMC SNR N49 \citep{2010A&A...518L.139O}. This structure was not seen in 4790\,MHz image reported by \citet{1993ApJ...411..761A}.  

%We currently have no reasonable explanation for the nature of this feature.

%(see Fig.~\ref{colored2}). 

%We detect a bridge like structure in 5500\,MHz image. This structure is consistence with both X-ray and optical images (see Figure~\ref{colored1}). 

%{\bf (REFERENCE; I think papers by Nigel, Milorad...)}

% \begin{figure*}
%     \centering
%     \includegraphics[scale=0.6, trim=0 0 0 0,clip]{RGB_final.eps}
%     %\includegraphics[scale=0.44, trim=0 0 0 0,angle=270,clip]{colerd_5500_new.eps}
%     %\includegraphics[scale=0.44, trim=0 40 0 0 0,angle=270,clip]{colored_9000_new.eps}
%     \caption{RGB image of E0102 SNR. R is ATCA image at 5500\,MHz frequency, G is HST image \citep{2000ApJ...537..667B}, and B is \cxo~ broadband image. The white plus represents the centre of expansion from \citet{2021AAS...2375...5120B}.}
%      \label{colored3}
% \end{figure*}

% \begin{figure*}
%     \centering
%     \includegraphics[scale=0.7, trim=0 0 0 0,clip]{new_1.eps}
%     %\includegraphics[scale=0.44, trim=0 0 0 0,angle=270,clip]{colerd_5500_new.eps}
%     %\includegraphics[scale=0.44, trim=0 40 0 0 0,angle=270,clip]{colored_9000_new.eps}
%     \caption{\emph{Upper panel:} ATCA image at 5500\,MHz of E0102 (left), \cxo\ broadband image (right). \emph{Lower panel:} HST image \citep{2000ApJ...537..667B} (left), and  RGB image (right). R is ATCA image at 5500\,MHz frequency, G is HST image \citep{2000ApJ...537..667B}, and B is \cxo~ broadband image.}
%      \label{colored4}
% \end{figure*}

%%%%%%%%%%%%%%%%%%%%%%%%%%%%%%%%%% Fig. 5 %%%%%%%%%%%%%%%%%%%%%%%%%%
 %The central white plus represents the centre of expansion as per  \citet{2021ApJ...912...33B}
%%%%%%%%%%%%%%%%%%%%%%%%%%%%%%%%%%%%%%%%%%%%%%%%%%%%%%%%%%%%%%%%%

\subsubsection{Central region}
Another interesting feature that we detect exclusively in our 9000\,MHz image is a horizontal bridge or bar-like feature in the central region with a measured flux density of 4.3 mJy. This feature is prominent across the interior of the SNR starting at the Eastern shell going almost all the way to the Western shell (see Fig.~\ref{colored4}, lower panel). The fact that this structure is observed across multiple array configurations with the ATCA, and on different dates, suggests this to be a real and interesting additional feature. This structure was not seen in the previous 8640\,MHz observations, likely due to lower sensitivity and poor $uv$ converge.
%We initially considered that this might be some sort of an observational artefact (due to calibration error or several imaging issues) but the fact that we see this feature at various arrays and across various observing dates argue for its true nature. 
%While we considered various explanation for this feature (including CCO, pulsar wind nebula (PWN), or jet like feature of a runaway pulsar), no firm conclusion could be reached as the resolution and sensitivity is somewhat poor. 
We are unable to conclusively deduce its true nature due to the limited sensitivity of our images at 9000\,MHz. Possible explanations may include a pulsar wind nebula (PWN) or jet-like features from a runaway pulsar. Further observations are required to draw any definitive conclusions.

\subsection{Spectral index}
 \label{SI}

The radio spectrum of an SNR can often be described as a pure power-law of frequency: $S_{\nu}$~$\propto$~$\nu^\alpha$, where $S_{\nu}$ is the flux density, $\nu$ is the frequency, and $\alpha$ is the spectral index.

We used the \textsc{miriad} \citep{1995ASPC...77..433S} task \textsc{imfit} to extract a total integrated flux density from all available radio continuum observations of SNR E0102 listed in Table~\ref{tab2}\footnote{Note that the flux density measurements for some images were also used in the \citet{2019AA} spectral index plot that is based on flux density measurements as explained in this section.}. 
This includes observations from the Murchison Widefield Array (MWA), MOST \citep{1976AuJPA..40....1C,1998PASA...15..280T}, ASKAP, ATCA \citep{2002MNRAS.335.1085F} and ALMA. We measured the MWA flux density for each sub-band (76--227\,MHz) and we also re-measured the ASKAP flux density at 1320\,MHz from \citet{2019MNRAS.490.1202J}. 
For cross-checking and consistency, we used \textsc{aegean} \citep{hancock2018} and found no significant difference in integrated flux density estimates. 
Namely, we measured E0102 local background noise (1$\sigma$) and carefully selected the exact area of the SNR which also excludes all obvious unrelated point sources. We then estimated the sum of all brightnesses above 5$\sigma$ of each individual pixel within that area and converted it to SNR integrated flux density following \citet[][eq.~24]{1966ARA&A...4...77F}. 
We also estimate that the corresponding radio flux density errors are below 10\,per\,cent as examined in our previous work \citep{2022MNRAS.512..265F,2023MNRAS.518.2574B}. In this estimate, various contributions to the flux density error are considered including missing short spacings. While for weaker sources this uncertainty is more proclaimed, for brighter objects such as our SNR E0102, the flux density error is much smaller ($<$10 per\,cent).

%We have also combined these total integrated flux density measurements from these two ATCA images with flux density measurements from the Murchison Widefield Array (MWA), MOST, ASKAP, ATCA, and ALMA. We measured the MWA flux density for each sub-band (76--227\,MHz) and we also re-measured the ASKAP flux density at 1320\,MHz from \citet{2019MNRAS.490.1202J}. 
%The MOST image at 408\,MHz was published in \citet{1976AuJPA..40....1C}, MOST image at 843\,MHz was published in \citet{1998PASA...15..280T}, and ATCA images at 2370, 4800, and 8640\,MHz were published in  \citet{2002MNRAS.335.1085F}. 
%Note that the flux density measurements for these images were also used in \citet{2019AA}  based on Filipovi\'c (priv. com.) measurements. 
%(see Table~\ref{tab2}). 

\begin{figure}
    \centering
    \includegraphics[scale=0.45, trim=0 0 0 0,clip]{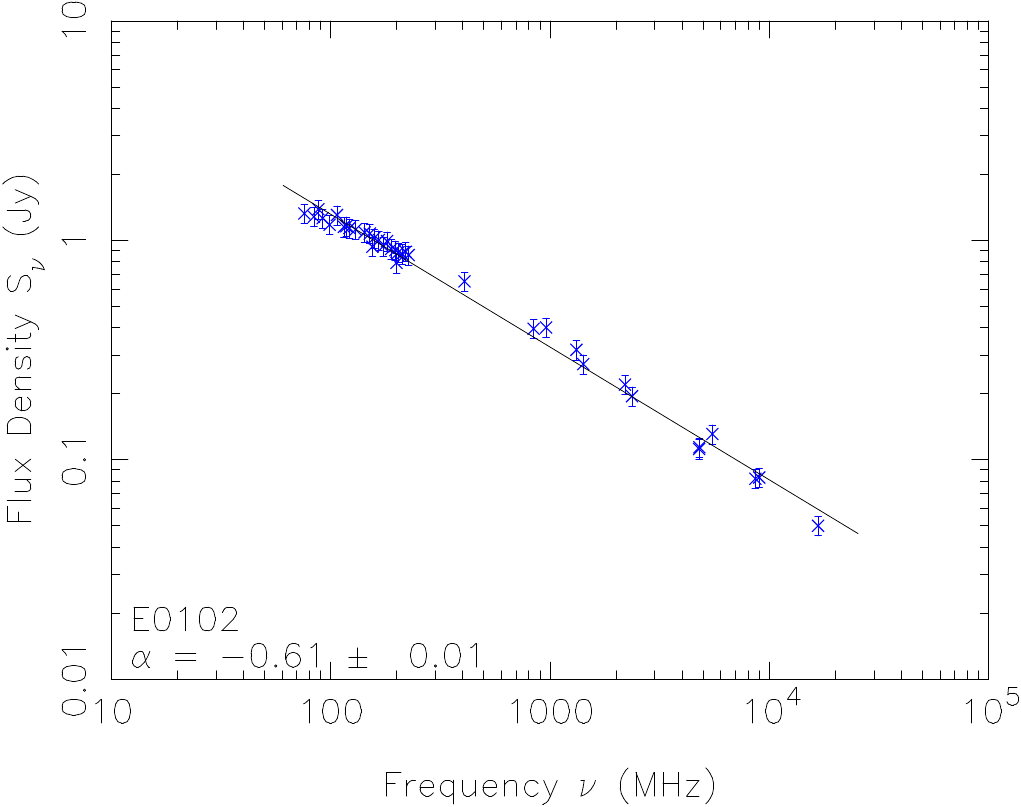}
    \caption{Radio continuum spectrum of E0102.}
    \label{fig:SI2}
\end{figure}

In Fig.~\ref{fig:SI2} we present the flux density vs.~frequency graph for E0102. The relative errors 
%(assumed 10\,per\,cent uncertainty in all flux density measurements) 
are used for the error bars on a logarithmic plot. The best power-law weighted least-squares fit is shown (thick black line), with the spatially integrated spectral index $\langle\alpha\rangle= -0.61 \pm 0.01$\footnote{We did not use ALMA flux in the fit as it is affected from missing short spacings.} which is somewhat flattened compared to the value of $\alpha=-0.7$ reported in the previous study of \citet{1993ApJ...411..761A}, and consistent with the value of the similar aged LMC SNR N\,132D of $-0.65\pm0.04$ \citep{2017ApJS..230....2B}. 

% I just cant find a meaning for flatter that makes any sense-it means to praise someone but to speak of slope, flatten is the root word to use.

%We fitted a two component spectral index, $\langle\alpha\rangle= -0.57 \pm 0.01$ (red line) and $\langle\alpha\rangle= -0.88 \pm 0.02$ (green line).

%We note that the spectral index becomes steeper towards high frequencies (108000\,MHz) suggesting a possible turn-over, though this is more likely to be because of the missing short spacings, as this effect is more pronounced at higher frequencies. 

%\Roland{I would say that this break in the spectrum is definitely caused by missing short spacings. In the 9000 MHz total power map in easily be seen that there are already short spacings missing as there area two areas, north and south of this central bar, where the flux goes significantly into negatives. And for ALMA this is a huge object.}
%\dots

%%%%%%%%%%%%%%%%%%%%%%%%%%%%%%%%%%% Table 2 %%%%%%%%%%%%%%%%%%%%%%%%
\begin{table}
	\centering
	\caption{E0102 flux density measurements used in this study (see Fig.~\ref{fig:SI2}). The asterisk ($^{*}$) indicates that we re-measured this flux density. 
 Note that the flux density values for the spectral index plot shown in \citep{2019AA} are based on measurements as explained in Section~\ref{SI}.
% Note that the flux density values for spectral index plot shown in \citep{2019AA} are based on measurements from Filipovi\'c (private communication).
	\label{tab2}}
	\begin{tabular}{@{}clcl@{}}
		\hline
$\nu$ &$S_{\nu}$&Telescope& Reference  \\
(MHz) & (Jy)   &          &        \\
		\hline
76    & 1.327  &MWA       & This work \\%\citet{2019AA}
84    & 1.281  &MWA       & This work \\%\citet{2019AA}
88    & 1.386  &MWA       & This work \\%\citet{2019AA}
92    & 1.261  &MWA       & This work \\%\citet{2019AA}
99    & 1.179  &MWA       & This work \\%\citet{2019AA}
107   & 1.305  &MWA       & This work \\%\citet{2019AA}
115   & 1.152  &MWA       & This work \\%\citet{2019AA}
118   & 1.161  &MWA       & This work \\%\citet{2019AA}
122   & 1.138  &MWA       & This work \\%\citet{2019AA}
130   & 1.121  &MWA       & This work \\%\citet{2019AA}
143   & 1.087  &MWA       & This work \\%\citet{2019AA}
151   & 1.074  &MWA       & This work \\%\citet{2019AA}
155   & 0.937  &MWA       & This work \\%\citet{2019AA}
158   & 1.026  &MWA       & This work \\%\citet{2019AA}
166   & 1.004  &MWA       & This work \\%\citet{2019AA}
174   & 0.939  &MWA       & This work \\%\citet{2019AA}
181   &0.993   &MWA       & This work \\%\citet{2019AA}
189   &0.915   &MWA       & This work \\%\citet{2019AA}
197   &0.903   &MWA       & This work \\%\citet{2019AA}
200   &0.790   &MWA       & This work \\%\citet{2019AA}
204   &0.885   &MWA       & This work \\%\citet{2019AA}
212   &0.868   &MWA       & This work \\%\citet{2019AA}
220   &0.888   &MWA       & This work \\%\citet{2019AA}
227   &0.857   &MWA       & This work \\%\citet{2019AA}
408   &0.650   &MOST      & \citet{2019AA} \\
843   &0.396   &MOST      & \citet{2019AA} \\
%900   &0.363   &MeerKAT   & Cotton et al. (in prep.)               \\
%950   &0.344   &MeerKAT   & Cotton et al. (in prep.)               \\
960   & 0.402   &ASKAP     & \citet{2019MNRAS.490.1202J}\\
%990   &0.338   &MeerKAT   & Cotton et al. (in prep.)               \\
%1034  &0.325   &MeerKAT   & Cotton et al. (in prep.)               \\
%1092  &0.313   &MeerKAT   & Cotton et al. (in prep.)               \\
%1144  &0.304   &MeerKAT   & Cotton et al. (in prep.)               \\
%1317  &0.273   &MeerKAT   & Cotton et al. (in prep.)               \\
1320  &0.317$^{*}$   &ASKAP     & \citet{2019MNRAS.490.1202J}\\
%1381  &0.268   &MeerKAT   & Cotton et al. (in prep.)               \\
1420  &0.273   &ATCA      & \citet{2019AA}\\
%1448  &0.264   &MeerKAT   & Cotton et al. (in prep.)               \\
%1519  &0.261   &MeerKAT   & Cotton et al. (in prep.)               \\
%1593  &0.238   &MeerKAT   & Cotton et al. (in prep.)               \\
%1656  &0.244   &MeerKAT   & Cotton et al. (in prep.)               \\
2100  & 0.220  &ATCA      &This work\\
2370  &0.195   &ATCA      &\citet{2019AA}\\
4790  &  0.112 &  ATCA     &\cite{1993ApJ...411..761A}\\
4800  &0.114   &ATCA      &\citet{2019AA}\\
5500  &0.131   &ATCA      & This work\\
8640  &0.082   &ATCA      & \citet{2019AA}\\
9000  &0.083   &ATCA      & This work\\
16700&0.050     &ATCA      &\citet{2019AA}\\
108000   &0.009   &ALMA      & This work\\
		\hline
	\end{tabular}
%		\begin{tablenotes}
%		 \item Column 1 is the observed frequency, column 2 the resulting flux density, column 3 the telescope used, and column 4 the paper that the measurement was taken from. MWA~=~Murchison Widefield Array, MOST~=~Molonglo Observatory Synthesis Telescope, ATCA~=~Australia Telescope Compact Array, and ASKAP~=~Australian Square Kilometre Array Pathfinder.
%		 \end{tablenotes}
\end{table}

%%%%%%%%%%%%%%%%%%%%%%%%%%%%%Fig 8 Two components %%%%%%%%%%%%%%%%%%%%
%%%%%%%%%%%%%%%%%%%%%%%%%%%%%%%%%%%%%%%%%%%%%%%%%%%%%%%%%%%%%%%%%%

\subsubsection{Spectral index image}

\begin{figure}
    \centering
    \includegraphics[ trim=0 0 0 0,angle=0,clip,scale=0.37]{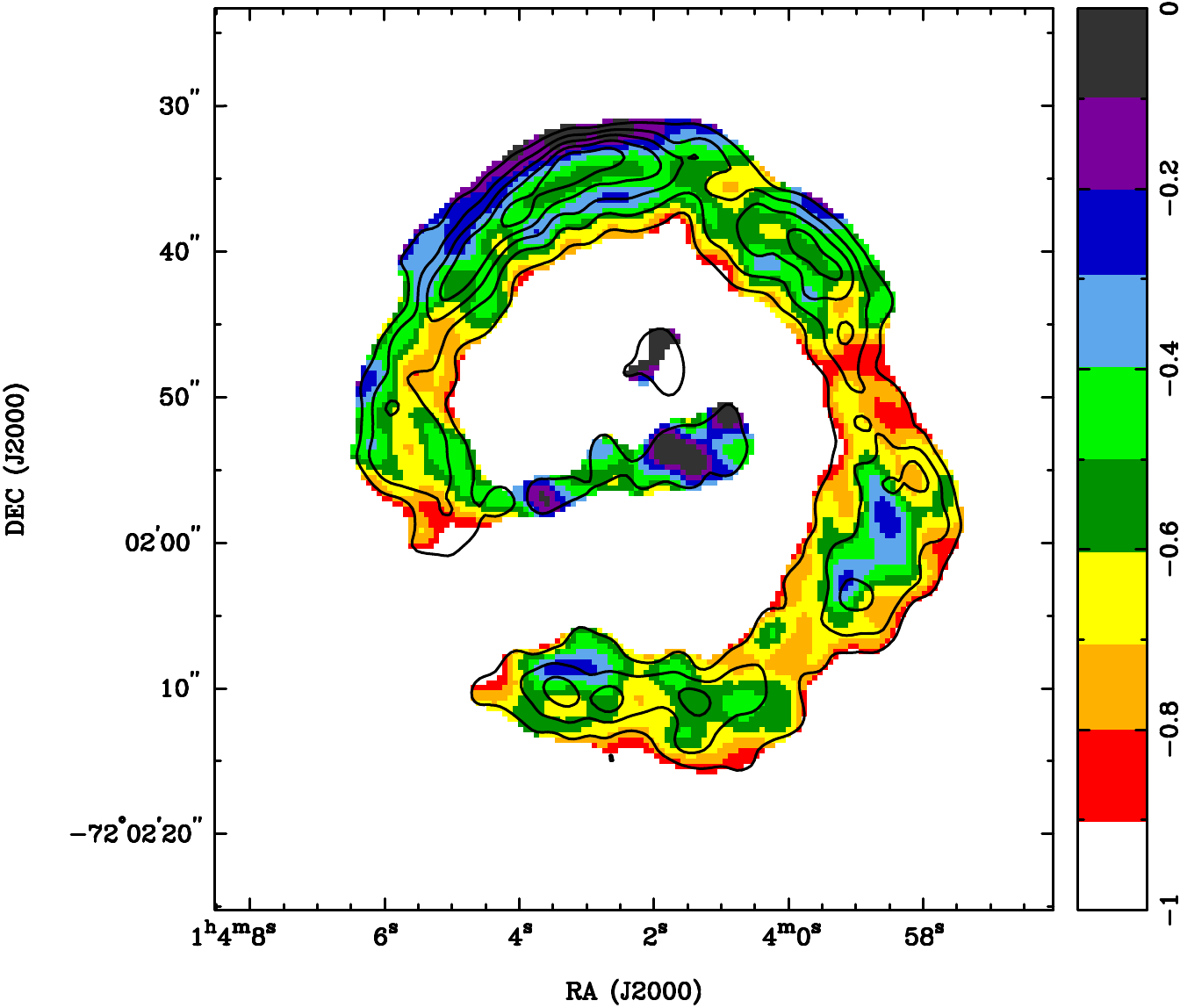}
    \caption{Spectral index map of E0102 (ATCA 2100, 5500, and 9000\,MHz images with 9000\,MHz contour lines overlaid). The contour levels are the same as Fig.~\ref{colored4}. The colour bar on the right-hand side represents the gradients of the spectral index. }
    \label{fig:SI1}
\end{figure}

We produced a spectral index map for E0102 using 2100, 5500 and 9000\,MHz images (Fig.~\ref{fig:SI1}). To do this, the images were re-gridded to the finest image pixel size ($0.3\times0.3$\,arcsec$^2$) using the \textsc{miriad} task \textsc{regrid}. These were smoothed to the lowest data resolution  ($2.7\times2.6$\,arcsec$^2$) using the \textsc{miriad} task \textsc{convol}. The \textsc{miriad} task \textsc{maths} then created the spectral index map (Fig.~\ref{fig:SI1}).

The mean value of the spectral index across the entire SNR is $-0.54\pm0.11$. 
%\Roland{How did you calculate the mean value? Is it the average or the median? Did you weigh individual pixels with their total power emission?} \Rami{I used the average value }
%from Denis
Most areas near the circumference (see Fig.~\ref{fig:SI1}) have a steep spectral index ($\alpha=-0.6$) at both inner and outer radii, while indices with flat gradients are found at intermediate radii ($\alpha=-0.1$). The north-east region is different in that there is no steepening at the largest radii but near zero gradients. 
The radio emission is brightest in the north-east with a close match between brighter emission and near zero spectral gradients. 
%The rotation measure (next section below) tends to be highest interior to the flattest region (Figure~\ref{fig:RM_boxes}) 
%Denis: need to replot figure 9 showing on the NE region, so the RM boxes are clearly visible. I think for the rest of the SNR the RM was not detectable (no measurements there).
The fractional polarisation (see Section~\ref{pol}) shown in Fig.~\ref{fig:pol.5500} is not well correlated with areas exhibiting steeper spectral indices.

We suggest that we are seeing a geometrical projection effect.
% the radio emission is coming from the forward shock region.%The X-ray emission mainly from higher density reverse shock region {\bf (see attached projected X-ray surface brightness-upper right panel for a CC model for E0102 that I just ran with ejecta mass 20~M$_{\odot}$ (forward shock is at normalized radius 1).}  
The radio emission comes from the forward shock region with higher energy electrons (with larger diffusion length) spread over a wider range in radius. This yields a flattened radio spectra further from the shock. 
The radio emission is also sensitive to the strength and orientation of the magnetic field, so SNRs tend to have patchy radio emission (more so than the X-ray emission). This is seen in radio images of most Galactic SNRs (e.g. see the radio images posted on SNRcat\footnote{\url{http://snrcat.physics.umanitoba.ca/d3/SNRmap/index.php}}) where they consist of partial shells or filaments.
%{\bf This is much like putting random blobs of paint on the surface of a balloon. }
Adding this patchiness at different radii, we see the whole thing in projection on the sky. If the centre of a bright patch happens to be on the projected SNR's rim, the flat spectral index region can be seen distinctly. But in most cases the bright patch, although on the surface of the 3D SNR, will not be on the rim so we see a combination of flat and steep spectral indices along the line of sight. For E0102 the brightest patch (north-east) happens to be nearly perfectly perpendicular to our line-of-sight, which would be a rare occurrence among SNRs. Finally, there are steeper spectral index regions ($-1.0<\alpha<-0.8$) towards the south-west and east SNR ring in the so-called SNR break-out regions similar to those seen in MCSNR\,0455–6838  \citep{2008SerAJ.177...61C}.

%While, the spectral index towards the northern and north-east parts of this remnant is gradually flattening from the middle of of radio emission where steep $\alpha$ of --0.6 is present to significantly flatter $\alpha=-0.1$ at the north-east edge. Completely opposite spectral index behavior occurs in the south and south-west parts (see Figure~\ref{fig:SI1}). 

%The northern and north-east regions are also most likely expanding into a rarefied environment, while emission in the south and south-west could be interacting with surrounding \HI\ emission (see Section~\ref{HI}). We see similar behaviour in G266.2--1.2 SNR \citep{2018ApJ...866...76M}.

%Finally, the spectral index in the central region is vary from steep ($\alpha=-0.68$) to flat ($\alpha=0.19$). 

%%%%%%%%%%%%%%%%%%%%%%%%% Fig. 7 %%%%%%%%%%%%%%%%%%%%%%%%%%%%%%%%%
%%%%%%%%%%%%%%%%%%%%%%%%%%%%%%%%%%%%%%%%%%%%%%%%%%%%%%%%%%%%%%%%%%%

%However, still inline with the values of other young SNRs in the LMC \citep{2017ApJS..230....2B}

\subsection{Polarisation}
\label{pol}

The fractional polarisation ($P$) can be calculated using the equation:

\begin{equation}
\label{eq}
P=\frac{\sqrt{S^2_Q+S^2_U}}{S_I},
\end{equation}

%\Roland{I hope you did not use this equation the way it is written here. For the mean fractional polarization, the polarized intensity map of the SNR, calculated from the Q and U maps, should be integrated and then divided by the integrated I map. Q or U integrated over the whole SNR should result in essentially 0. It would also be interesting to note peak fractional polarizations or the percentage polarization at selected location. For example the percentage polarization at the centre of the bright north-eastern shell is about 18\% at 5500 MHz and 23\% at 9000 MHz. }

%\Rami{I didn't use equation 1, I just put it as general equation.}

\noindent where $P$ is the mean fractional polarisation; $S_{Q}$, $S_{U}$, and $S_{I}$ are %integrated \Roland{as I said above, remove the word integrated as it implies that you summed up several values of Q and U, in which case negative and positive values would average out!}
intensities for the $Q$, $U$, and $I$ Stoke parameters, respectively. 

\begin{figure*}
\centering
%\hspace{-0.42\textwidth}a)\hspace{0.47\textwidth}b) \\
\includegraphics[width=0.43\textwidth,angle=0,trim=70 45 0 0,clip]{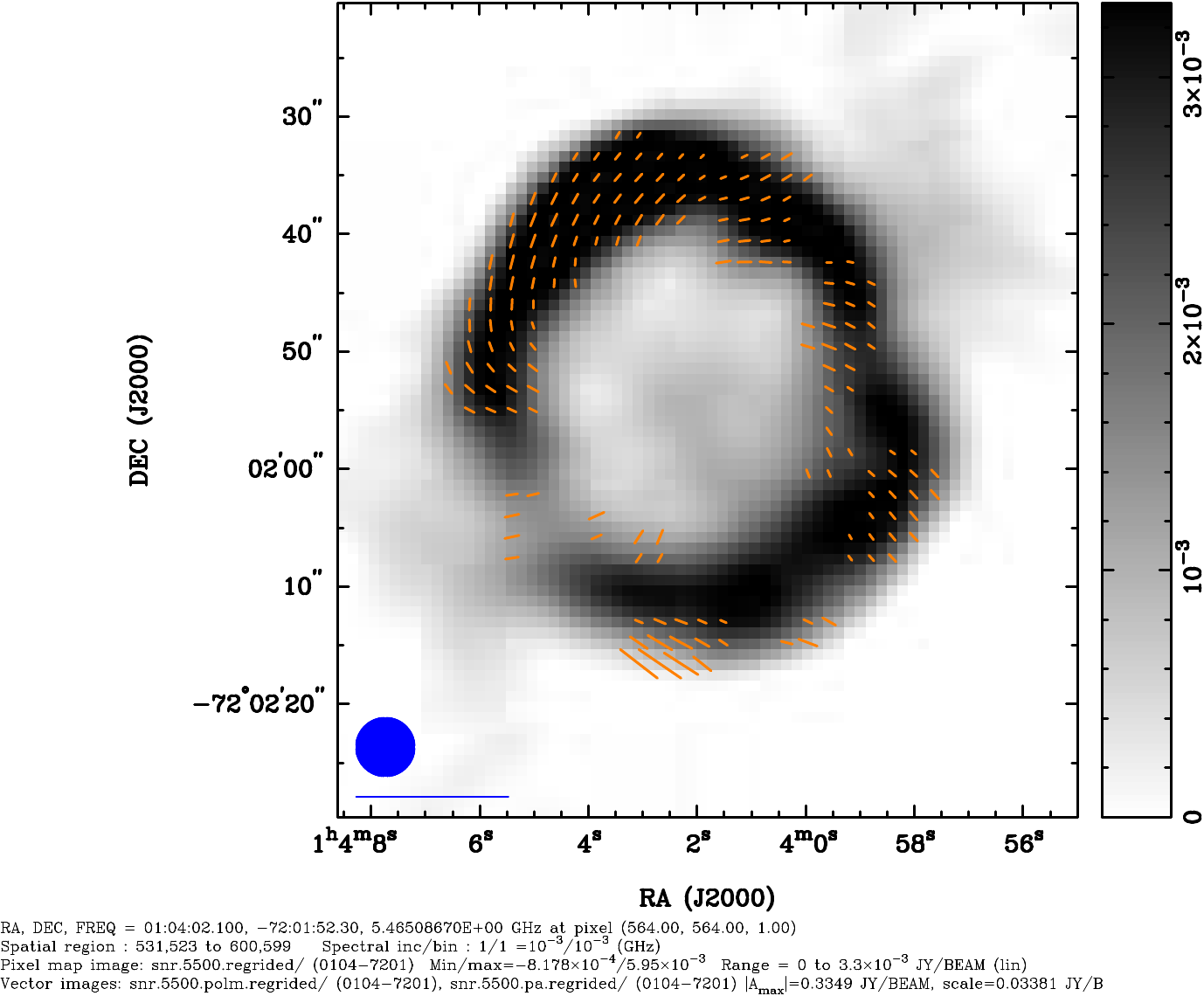}
\includegraphics[width=0.43\textwidth,angle=0,trim=70 45 0 0,clip]{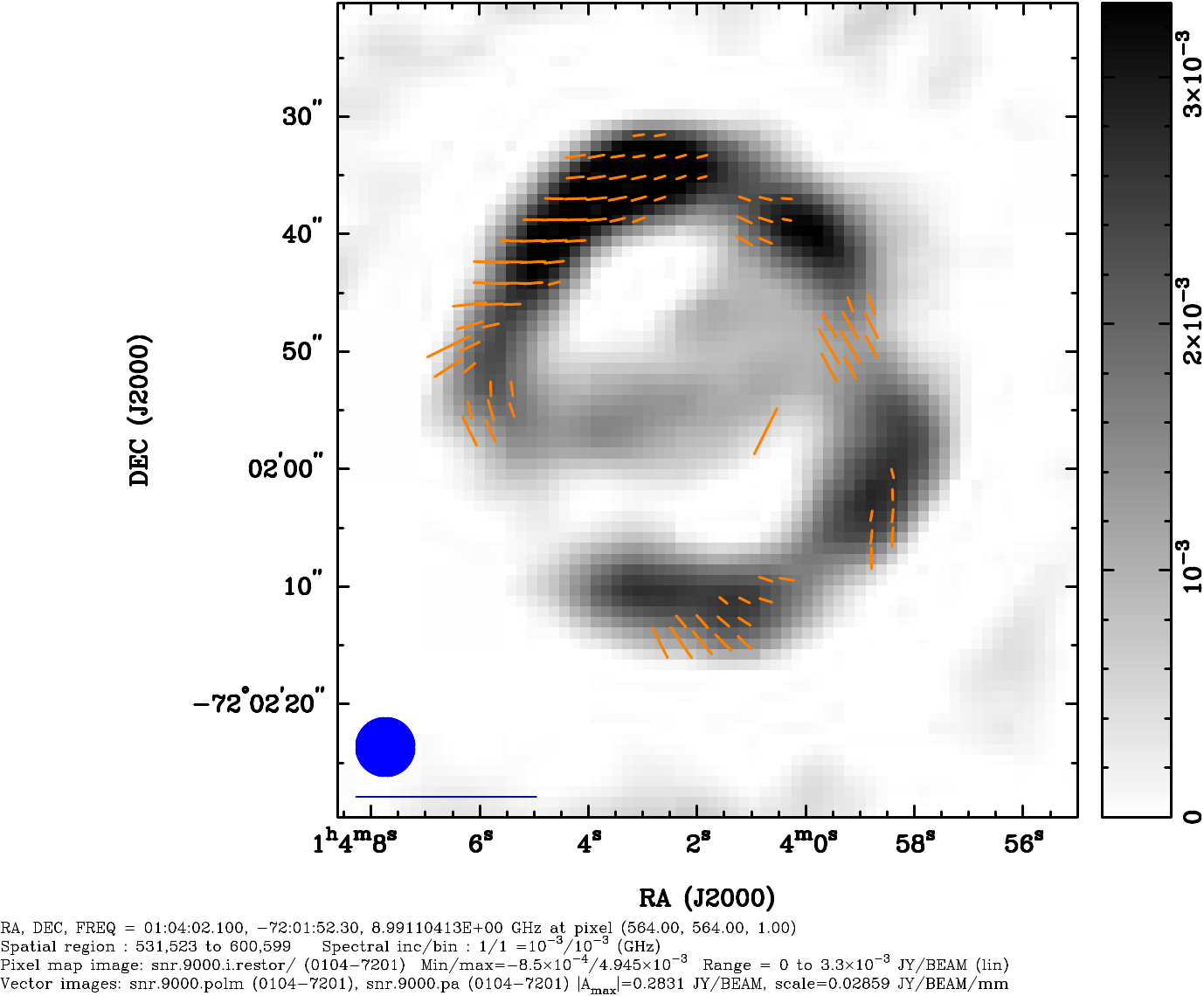}\\
%\hspace{-0.42\textwidth}c)\hspace{0.47\textwidth}d) \\
\includegraphics[width=0.43\textwidth, trim=0 0 0 0,angle=0,clip]{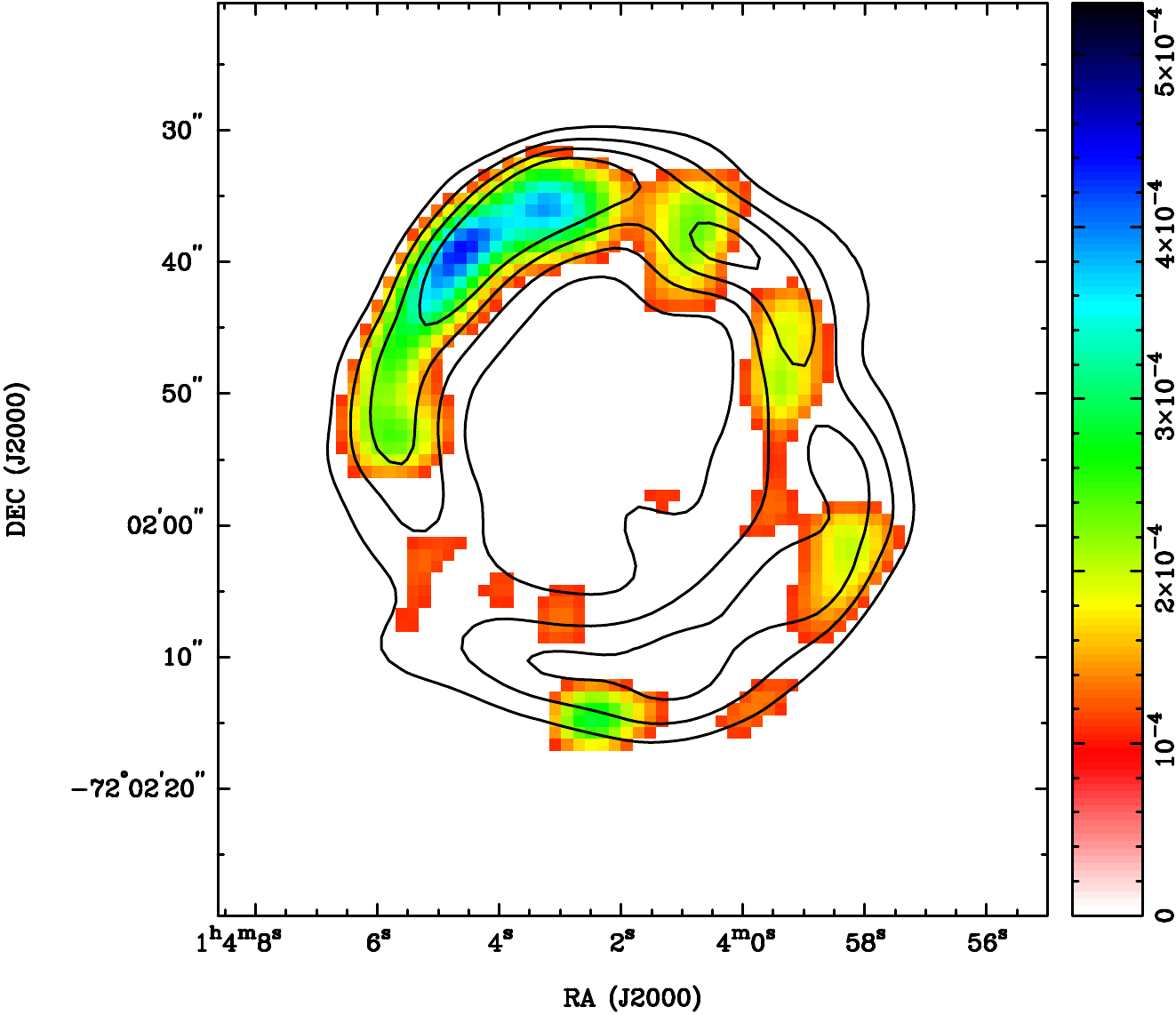}
\includegraphics[width=0.43\textwidth, trim=0 0 0 0,angle=0,clip]{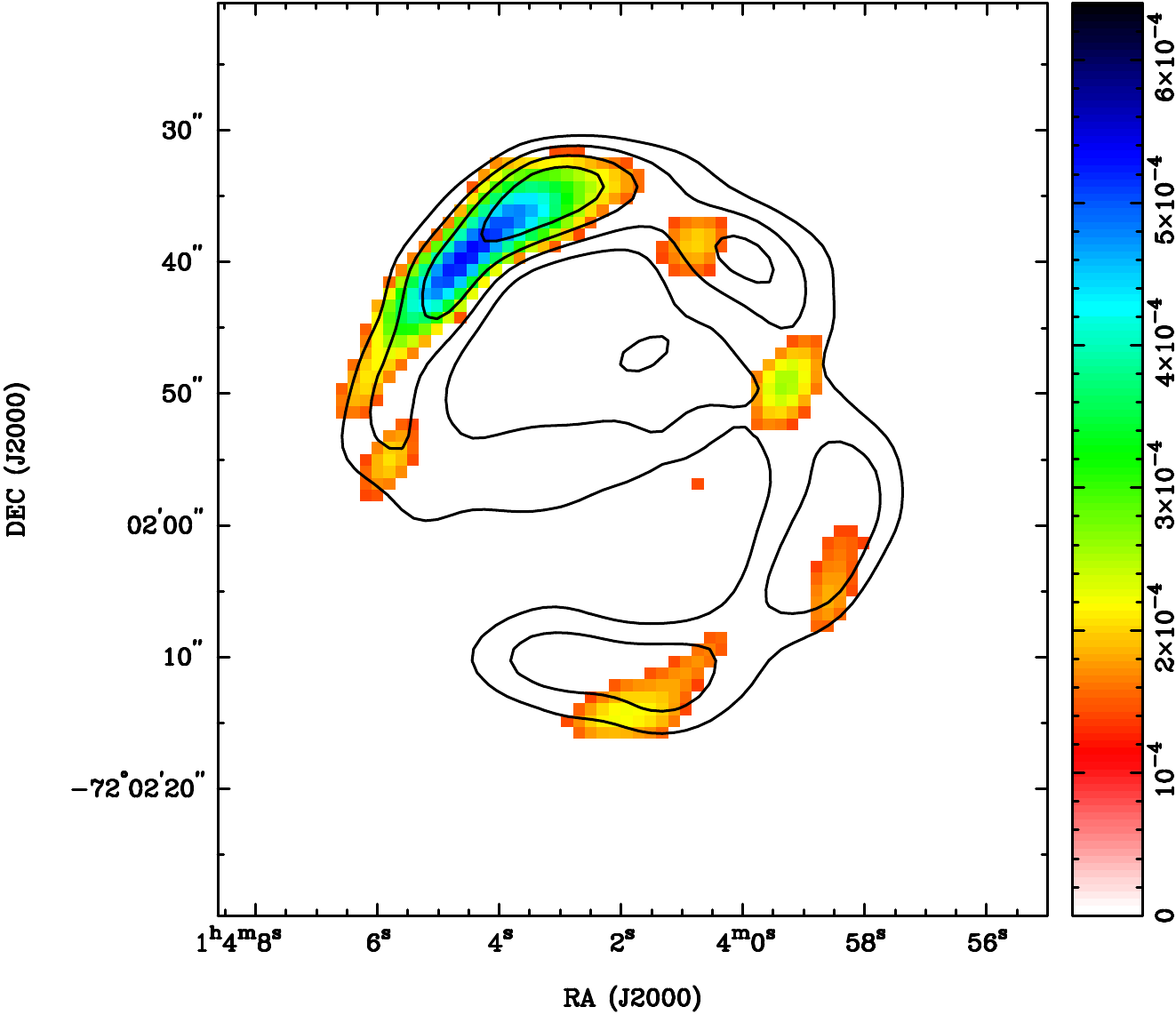}\\
\caption{Fractional polarisation vectors of E0102 overlaid on intensity ATCA image at 5500\,MHz (upper left) and at 9000\,MHz (upper right). The blue circle in the lower left corner represent a synthesised beam of $5\times5$\,arcsec$^2$ and the blue line below the circle represents 100\,per\,cent polarisation. The bar on the right side represents the grayscale gradients for the ATCA images in Jy\,beam$^{-1}$. Polarisation intensity maps of E0102 at 5500\,MHz (bottom left) and at 9000\,MHz (bottom right) with intensity image contour lines overlaid. The contour levels are 0.001, 0.002, 0.003, and 0.004\,Jy\,beam$^{-1}$ for both images. The colour bar represents gradients of polarisation intensity in Jy\,beam$^{-1}$.} %c) Fractional polarisation vectors of E0102 overlaid on intensity ATCA image at 9000\,MHz. The filled and open blue circles in the lower left corner represents a synthesised beam of $2\times2$ and $5\times5$\,arcsec, respectively; and the blue line below the circle represents 100\,per\,cent polarisation. d) Polarisation intensity map of E0102 at 9000\,MHz with intensity image contour lines overlaid. The contour levels are as the same as Fig.~\ref{colored2}. The colour bar represents gradients of polarisation intensity}
\label{fig:pol.5500}
\end{figure*}

We used the \textsc{miriad} task \textsc{impol} to produce polarisation maps. The fractional polarisation from E0102 appears prominent at both 5500\,MHz and 9000\,MHz towards the north-east and north-west of the SNR, due to interactions with the local ISM. The west and south-west remnant both show slight polarisation. The polarisation vector maps at both  5500\,MHz and 9000\,MHz are presented in Fig.~\ref{fig:pol.5500} as well as the polarisation intensity maps at each frequency.

We estimate a mean fractional polarisation  of $7\pm1$\,per\,cent at 5500\,MHz and $12\pm2$\,per\,cent at 9000\,MHz, with a peak value at the centre of the bright north-eastern shell of $9\pm1$\,per\,cent and $26\pm6$\,per\,cent, respectively. The fractional polarisation of E0102 is somewhat higher than the 4\,per\,cent (4790\,MHz) \citep{1995AJ....109..200D} value of similar age \citep[$\sim$2500\,yr;][]{2020ApJ...894...73L} oxygen-rich LMC SNR N\,132D. 
%The mean values of $P$ towards these regions are $14\pm3$\,per\,cent and $17\pm 3$\,per\,cent for 5500 and 9000\,MHz, respectively. 
We did not detect any significant polarisation towards the bridge-like structure and the central region, with upper limits of 2 and 4\,per\,sent for 5500 and 9000\,MHz, respectively. This lack of polarisation argues against a PWN scenario for this emission, as PWNs are typically associated with strong polarisation \citep[e.g.,][]{2012A&A...543A.154H, 2014ApJ...780...50B}.

%%%%%%%%%%%%%%%%%%%%%%%%%%%%%%%%%%%%%%%%%%%%%%%%%%%%%%%%%%%%%%%%%
% \begin{figure}
% \centering
% \includegraphics[scale=0.43,angle=270,trim=0 70 45 0,clip]{p_9000.eps}
% \caption{Fractional Polarisation vectors of E0102 overlaid on intensity ATCA image at 9000\,MHz, }
% \label{fig:pol.9000}
% \end{figure}

% \begin{figure}
%     \centering
%     \includegraphics[scale=0.38, trim=0 0 0 0,angle=270,clip]{PI_5500.eps}
%     \caption{Polarisation intensity map of E0102 at 5500\,MHz with intensity image contours overlaid. The contour levels are 0.0002, 0.0004, 0.0006, 0.0008, and 0.001\,Jy\,beam$^{-1}$.}
%     \label{fig:PI.5500}
% \end{figure}

% \begin{figure}
%     \centering
%     \includegraphics[scale=0.38, trim=0 0 0 0,angle=270,clip]{PI_9000.eps}
%     \caption{Polarisation intensity map of E0102 at 9000\,MHz with intensity image contours overlaid. The contour levels are 0.0002, 0.0004, 0.0006, 0.0008, and 0.001\,Jy\,beam$^{-1}$.}
%     \label{fig:PI.9000}
% \end{figure}

% \begin{figure*}
%     \centering
%     \includegraphics[scale=0.39, trim=0 0 0 0,clip]{three_final1.eps}
%     \caption{polarisation intensity maps of 0102 at (a) 9512\,MHz, (b) 9000\,MHz, and (c) 8488\,MHz. The four circles are the same as Figure~\ref{colored2}. }
%     \label{fig:three}
% \end{figure*}

% \begin{figure}
%     \centering
%     \includegraphics[scale=0.34, trim=0 0 0 0,angle=270,clip]{RM.eps}
%     \caption{Rotation measurement map of E0102 }
%     \label{fig:RM}
% \end{figure}

\subsection{Rotation measure}
\label{RM}

%To calculate the rotation measure ($RM$) of E0102, we split the 2048\,MHz bandwidth at 5500\,MHz into four 512\,MHz sub bands (4723, 5244, 5756, and 6268\,MHz)
One way to estimate the line-of-sight magnetic field in the direction of E0102 is by finding its rotation measure ($RM$) using $Q$ and $U$ maps at 5500 and 9000\,MHz, convolved to a common resolution of 4\,arcsec. 

The rotation measure is defined as the slope of the observed position angle ($\phi$) as a function of wavelength ($\lambda$) squared:

\begin{equation}
RM = \frac{d\phi(\lambda^2)}{d\lambda^2},
\label{eq2.1}
\end{equation}

\noindent To calculate the $RM$ and intrinsic polarisation angle maps we used the program \textsc{rmcalc} from the \textsc{DRAO} export software package \citep{1997ASPC..125...58H}.

 \begin{figure*}
     \centering
     \includegraphics[width=0.95\textwidth]{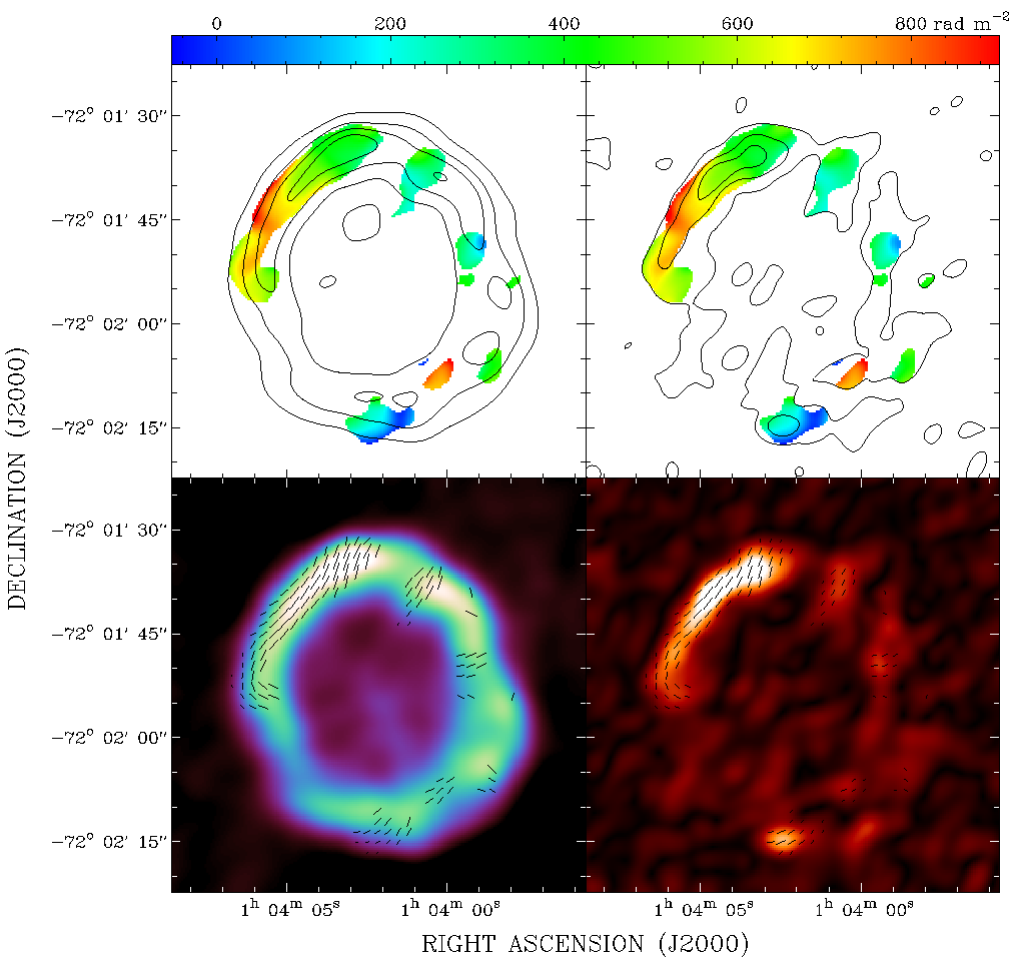}
     \caption{Top: $RM$ map calculated between the observations at 5500\,MHz and 9000\,MHz with overlaid contours of total power (left) and polarised intensity (right) at 5500\,MHz. Bottom: Images of total power (left) and polarised intensity (right) with overlaid vectors in B-field direction, corrected for Faraday rotation.}
     \label{fig:rm+phi0}
 \end{figure*}

The final $RM$ map is shown in Fig.~\ref{fig:rm+phi0}. We only calculated $RM$ where the signal-to-noise ratio in the polarised intensity maps was at least $10\sigma$ at both frequencies. We notice that all $RM$ values are positive, with a maximum of almost $1000\pm25$\,rad\,m$^{-2}$ at the outer edge of the bright north-eastern shell and a minimum of about 0\,rad\,m$^{-2}$ at the bottom of the SNR. One striking feature is a $RM$ gradient along the bright shell in the north-east, which could indicate the expansion into a homogenous interstellar magnetic field  \citep{2009IAUS..259...75K}. The foreground contribution of the MW is about +23\,rad\,m$^{-2}$ according to \citet{2008ApJ...688.1029M}, based on a $RM$ study of polarised background sources that are behind the SMC. Contributions from inside the SMC are not clear as the latest study of the SMC's internal magnetic field structure \citep{2022MNRAS.510..260L} did not have any samples in the area around E0102. They found slightly more negative than positive $RM$s, and most of those sources are in positive and negative clusters.

Once the $RM$ has been determined, we next rotated the polarisation vectors back to their intrinsic (zero wavelength) values. In Fig.~\ref{fig:rm+phi0} we have added an additional 90$\degr$ to the intrinsic electric vectors, so the vectors represent the magnetic-field directions in the remnant projected to the plane of the sky. In the shell of E0102, we find segments with tangential and radial magnetic fields, representing later and earlier phases of evolution, respectively. Most striking is the tangential magnetic field related to the prominent north-eastern part of the SNR's shell, where we also found the $RM$ gradient. 

\subsection{Magnetic fields in E0102}
\label{magneticfields}

Observed rotation measure can help  better understand the line-of-sight magnetic field in the direction of E0102.  Useful to this discussion is first exploring an independent method known as equipartition or minimum-energy. This method uses modeling and simple parameters to estimate intrinsic magnetic field strength and energy contained in the magnetic field and cosmic ray particles using radio synchrotron emission\footnote{\url{http://poincare.matf.bg.ac.rs/~arbo/eqp}} \citep{2012ApJ...746...79A,2013ApJ...777...31A,2018ApJ...855...59U}.% to estimate an average magnetic field strength for E0102. 

This approach is a purely analytical method, described as an order of magnitude estimate because of errors in determination of distance, angular diameter, spectral index, a filling factor, and flux density, tailored especially for the  magnetic field strength in SNRs.  \citet{2012ApJ...746...79A,2013ApJ...777...31A,2018ApJ...855...59U} present two models; the difference is composition of assumed cosmic ray particles. In their original "$\kappa \neq0$" model, where $\kappa$ is a constant in the power-law energy distributions for electrons, cosmic rays are assumed to be composed of electrons, protons and ions. Their "$\kappa=0$" model assume cosmic rays composed of electrons only. \citet{2018ApJ...855...59U} showed the latter type of equipartition is superior to the former.

%\textbf{If $\kappa=0$ we assume equipartition between cosmic ray electrons and magnetic field. \citet{2018ApJ...855...59U} showed that this type of equipartition is better for using than equipartition between all cosmic ray ingredients (electron, protons, and iones) with magnetic fields ($\kappa \neq0$)}.

Using the \citet{2018ApJ...855...59U} model\footnote{We use: $\alpha=0.61$, $\theta$~=~0.36\,arcmin, $\kappa=0$, $S_{\rm 5500\,MHz}$~=~0.131\,Jy, and f~=~0.84.}, the mean equipartition field over the whole E0102 remnant is $65\pm5$\,$\mu$G, with an estimated minimum energy of $E_{\rm min}=1\times10^{49}$\,erg. (The original model, cited in \citet{2012ApJ...746...79A}, yields a mean equipartition field of $150\pm5$\,$\mu$G, with an estimated minimum energy of $E_{\rm min}=5\times10^{49}$\,erg.) 
%\textbf{If $\kappa=0$ we assume equipartition between cosmic ray electrons and magnetic field. \citet{2018ApJ...855...59U} showed that this type of equipartition is better for using than equipartition between all cosmic ray ingredients (electron, protons, and iones) with magnetic fields ($\kappa \neq0$)}.

%These values indicate an amplified magnetic field produced by non-linear particle acceleration effects typical for young SNRs

%which is most likely an overestimate, because physical background gives better equipartition arguments for $\kappa=0$ \citep{2018ApJ...855...59U}. 
%start here
  To estimate the line-of-sight magnetic field, we use rotation measure and assume the polarisation angle changes linearly as a function of wavelength squared. The following equation may be used to estimate the magnetic field of E0102\footnote{Following \citet{1966MNRAS.133...67B}, this equation is known as the Faraday depth. If we assume there is only one source along the line of sight, which has no internal Faraday rotation, the Faraday depth approximates the rotation measure at all wavelengths \citep{2005A&A...441.1217B}.}:
%using our observed rotation measure and depolarisation of the radio shells:

\begin{equation}
RM=0.81\int_{L}^{0}n_{e}{\bf B}\cdot {\bf dl} = 0.81~\int_{L}^{0}n_{e}B_{\|}dl~~[{\rm rad\,m^2}],  
\label{eq2}    
\end{equation}
\noindent where $n_{e}$ is the electron density in cm$^{-3}$, $B_{\|}$ is the line-of-sight magnetic field strength in $\mu$G, and $dl$ represents the path length along  the Faraday rotating medium in parsecs. 

We concentrate our study on the bright shell in the north-east (top left) of the SNR as it seems %to have the most stable conditions and also seems 
to be the most evolved. The magnetic field, corrected for Faraday rotation, is tangential to the shell. This is an indication that the swept-up ambient magnetic field is dominating the uniform magnetic field and synchrotron emission, which also indicates the swept-up material is dominating the hydrodynamics and may have just entered the Sedov phase. We determined (Section~\ref{centre}) that the radius of the SNR in this region is 5.8\,pc, the lowest value; another indicator this area is most developed. To determine the compression ratio of the SNR in this region, we estimated the radial distance from the radio ridge to the edge of the SNR. Using this and the radius, we measure a compression ratio of 3.7, supporting the Sedov phase since our compression ratio estimate is near 4 \citep{1959sdmm.book.....S}. %Nevertheless, we proceed to use this compression ratio.  not sure what 'this' is??

 \begin{figure}
     \centering
     \includegraphics[width=0.45\textwidth]{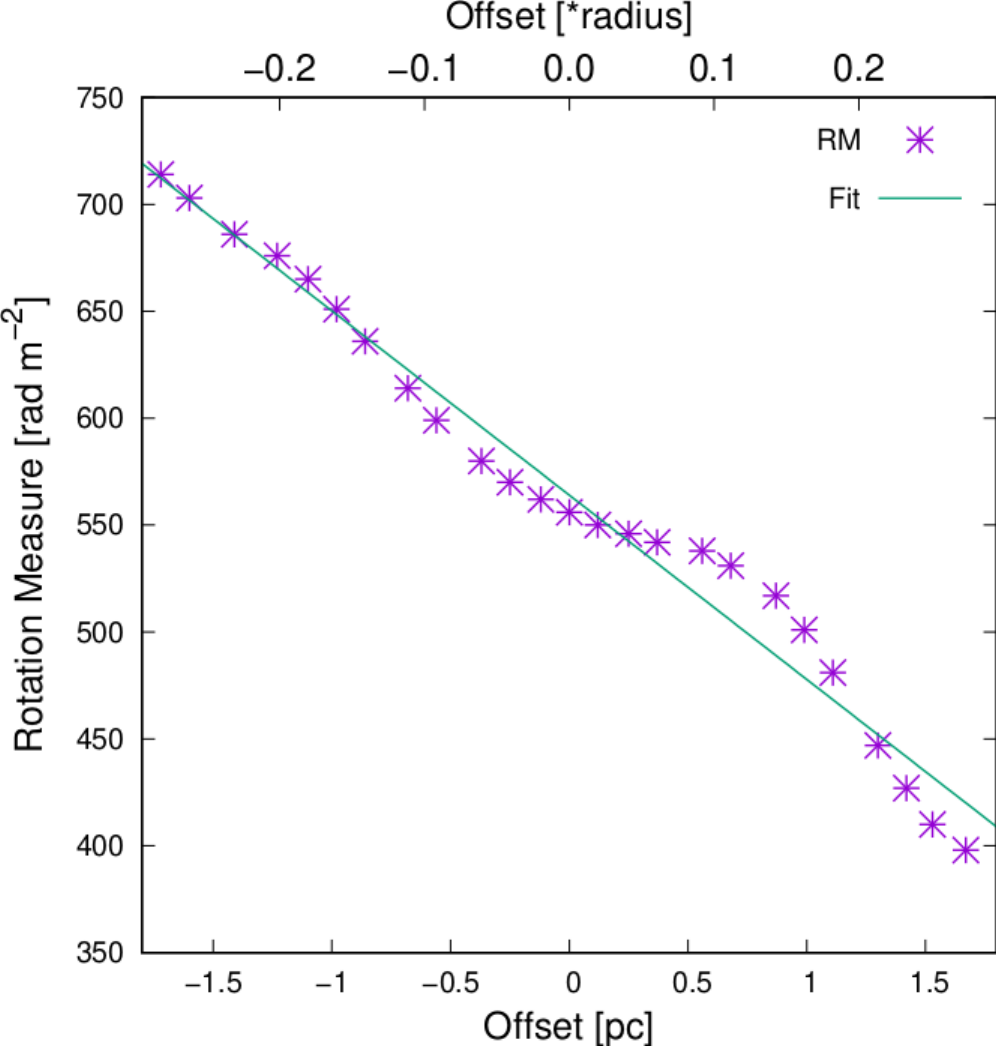}
     \caption{Rotation measure along the ridge of the bright shell in the north-east. The centre is at the peak in polarised intensity. The negative offset is to the lower left and the positive offset is to the upper right. The offset is given in pc (bottom) and in a fraction of the SNR's radius (top). The green line indicates a linear fit to the gradient.}
     \label{fig:rmridge}
 \end{figure}

%We plotted the rotation measure along the bright polarised intensity ridge of the north-east shell in Fig.~\ref{fig:rmridge}.
\citet{2009IAUS..259...75K} modelled radio polarisation observations of an SNR expanding inside a homogenous medium. We plotted our rotation measure along the bright polarised intensity ridge of the north-east shell in Fig.~\ref{fig:rmridge}. The rotation measure observed from the SNR's radio shell is dominated by the magnetic field in the nearby part of the SNR.  % as Faraday rotates most of the emission, and 
In this nearby part, one side the magnetic field is coming towards us and  the other is pointing away from us. If the ambient magnetic field is only in the plane of the sky (i.e., has no line-of-sight component) we should find a $RM$ of 0 in the centre, negative on one side and positive on the other. The centre $RM$ can be affected by the 
foreground $RM$ ($\sim$ +23\,rad\,m$^{-2}$ as noted in Section~\ref{RM}), the angle between the plane of the sky, and the ambient magnetic field. In this case, we can determine that the ambient uniform magnetic field is pointing towards us, and is pointing from the bottom left to the top right as the $RM$ is positive but decreasing along the same direction. It only requires a small rotation angle to have all $RM$s in the positive or the negative on the shell. We fitted the gradient with a linear function, also shown in Fig.~\ref{fig:rmridge}. The resulting $RM$ in the centre is $+531\pm5$\,rad\,m$^{-2}$, corrected for the foreground contribution of the MW. The contribution of the SMC is unclear as we stated in Section~\ref{RM}. At the bright ridge on the SNR's shell, located at the inner edge, the line-of-sight through the emission region is the longest. With a radius of 5.8\,pc and a compression ratio of 3.7, this line-of-sight should be 5.3\,pc. In the shell of the SNR, synchrotron emitting and Faraday rotating plasmas are intermixed. Therefore, the synchrotron emission generated at the back of the SNR sees the entire $RM$ and the emission generated at the front sees no $RM$. If all the parameters, as synchrotron emissivity, magnetic field strength and angle, and electron density are approximately constant along the line of sight through the SNR shell and the polarised emission is Faraday thin, the $RM$ seen by a background source is twice the value observed \citep{1966MNRAS.133...67B}. Using equation~\ref{eq2}, with $RM=2\times 531$\,rad\,m$^{-2} = 1062$\,rad\,m$^{-2}$, $dl$~=~5.3\,pc, and electron density $n_e$~=~14.8\,cm$^{-3}$ for the post-shock gas \citep{2017A&A...602L...4V}, the magnetic field strength of the uniform component inside the shell parallel to the line-of-sight is $B_\parallel$~=~16.7\,$\mu$G. 

We can assume the intrinsic percentage polarisation of the SNR's synchrotron emission through the spectral index to be 71\,per\,cent \citep{1961AnAp...24...71L}. In the shell of the SNR, where the synchrotron emitting and Faraday rotating plasmas are intermixed, a high rotation measure (as we measured) causes depolarisation even at such a high frequency. According to \citet{1998MNRAS.299..189S}, the $RM$ value we found in the middle of the shell, would reduce the percentage polarisation from 71\,per\,cent to 55.7\,per\,cent. But, in the centre of the shell, we observe a 15\,per\,cent linear polarisation. This additional depolarisation is caused by random fluctuations of the magnetic field \citep{1966MNRAS.133...67B}. From this depolarisation, we estimate the ratio between the uniform and random components of the magnetic field. This high depolarisation here requires a random magnetic field component of 27.5\,$\mu$G, using the above calculated 16.7\,$\mu$G as the uniform component. In this case, we would have a total magnetic field in the shell of $\sim$44\,$\mu$G. The uncertainty in magnetic field strength remains undetermined, primarily due to it being a combination of the uncertainties associated with the \textit{RM} and electron density, with the electron density's uncertainty being unquantified.

%\textbf{Uncertainties of the magnetic field should be a combination of the uncertainties of the \textit{RM} and the electron density.}  

This of course only takes the line-of-sight component of the magnetic field into account. To achieve the magnetic field strengths estimated through equipartition, we require an angle of 38\degr\ between the plane of the sky and the ambient magnetic field for $\sim65$\,$\mu$G\footnote{For the original eqipartition model, an angle of 17\degr, given a field strength of $\sim{150}$\,$\mu$G, would be required extrapolating to a total ambient magnetic field strength of 10.8\,$\mu$G when correcting for compression.}, containing a uniform magnetic field of 24.7\,$\mu$G, after scaling 16.7\,$\mu$G as given above. Correcting for compression, we can extrapolate a total ambient magnetic field strength of the uniform components to 5.1\,$\mu$G. Here, we cannot simply apply the linear compression ratio of 3.7, as the magnetic field is not only compressed but also enhanced. The `magnetic field lines' are frozen into the plasma of the expanding shell and therefore also stretched out around the perimeter of the SNR, which increases the magnetic energy \citep{1968ApJ...154..807W}. This leads to a combined compression and enhancement of 4.8.

%Correcting for compression, we can extrapolate a total ambient magnetic field strength of the uniform components to 5.1\,$\mu$G . \textcolor{red}{NOTE: Roland,we should explain how we correct for compression to get the 5.1 value, there must be a model or equation we are using.  Also we need to know significant value for these, thanks.}
%and 10.8\,$\mu$G, respectively.

\subsection{  \texorpdfstring{\HI}{H}/CO morphology} 
 \label{HI}

\begin{figure*}
    \centering
    \includegraphics[ trim=0 0 0 0,scale=0.9,clip]{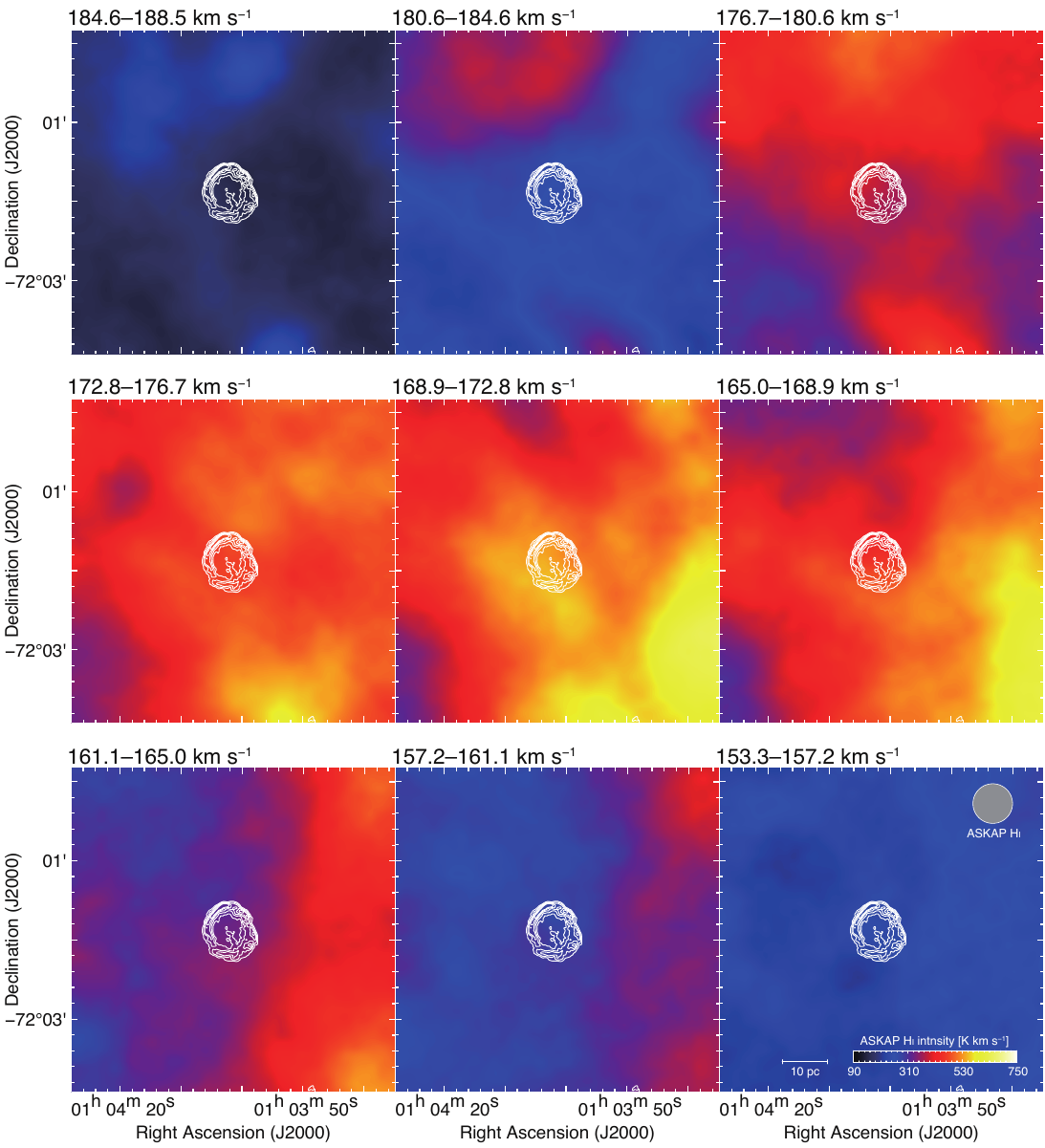}
    \caption{Velocity channel map of \HI\ towards E0102 obtained from the ASKAP \citep{2022PASA...39....5P}. Each panel shows the \HI\ intensity distribution integrated every 3.9\,km\,s$^{-1}$ in a velocity range from 188.5 to 153.3\,km\,s$^{-1}$. Our 5500\,MHz ATCA image has been overlaid on top with contours levels as per Fig.~\ref{colored4}. The grey circle and colour bar on the bottom right panel represent a synthesised beam of $30\times30$\,arcsec$^{2}$ and  gradients of \HI\ in K\,km\,s$^{-1}$, respectively. A bar for physical size on the sky is shown.}
    \label{fig:HI}
\end{figure*}

\begin{figure}
    \centering
    \includegraphics[width=\linewidth,clip]{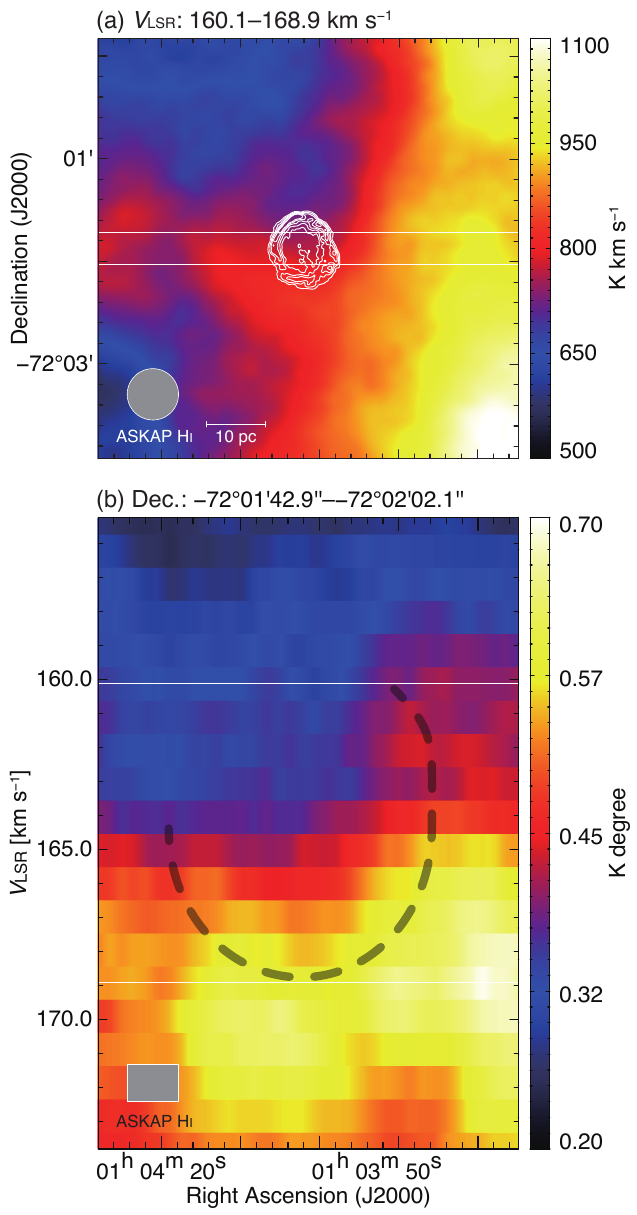}
    \caption{(a) \HI\ integrated velocity intensity map towards E0102, where the integration velocity range is from 160.1 to 168.9\,km\,s$^{-1}$. Our 5500\,MHz image is overlaid with intensity levels as per Fig.~\ref{colored4}. (b) Position--velocity diagram of \HI\ with Declination $-72^{\circ}01'42\farcs9$ to $-72^{\circ}02'02\farcs1$. The dashed curve delineates a possible expanding gas motion (see text).}
    \label{fig:pv}
\end{figure}

The E0102 SNR region was mapped in \HI\ using ASKAP from \cite{2022PASA...39....5P}; these velocity channel maps are shown in Fig.~\ref{fig:HI}, with our 5500\,MHz image overlaid.
%Fig. \ref{fig:HI} shows the velocity channel maps of \HI\ obtained with ASKAP \citep{2018NatAs...2..901M,2019MNRAS.483..392D}. 
\HI\ emission is visible in the velocity range of $\sim$160--180\,\kms. There 
appears to be a cavity-like structure forming near the location of E0102 in the velocity range 165.0--168.9\,\kms (see Fig.~\ref{fig:pv}, a). 
A large cloud is seen at $\sim$170--180\,\kms\ in the north at Dec~$>-$72$^\circ$02$\arcmin$, which then transitions to a cloud in the west at $\sim$160--175\,\kms (RA~$<$~01$^{h}$04$^{m}$). Those emission regions are most likely connected at the velocities around E0102. The boundary of \HI\ appears along the radio shell, especially towards the south-west. 
% and it seems to be that E0102 is sitting right in the region where the clouds overlap at $\sim$165\,\kms. 
%In the direction of E0102, we find a cavity-like structure of \HI\ at the velocity of 163.7--167.6\,km\,s$^{-1}$. 

The \HI\ distribution in the velocity space has unique features. Fig.~\ref{fig:pv}(b) shows the position--velocity ($p$--$v$) diagram of \HI. We found a cavity-like structure of \HI, whose size is roughly three times larger than the diameter of the SNR shell. We propose a possible scenario that the cavity-like structures in both the spatial distribution and $p$--$v$ diagram correspond to a wind-blown bubble that was formed by strong stellar winds from a high-mass progenitor of E0102 and is now partially interacting with an inner wall of the \HI\ cavity. The observational signature is similar to \HI\ studies of core-collapse SNRs in the SMC (DEM\,S5, \citealt{2019MNRAS.486.2507A}; RX\,J0046.5$-$7308, \citealt{2019ApJ...881...85S}). However, more data are required to draw definitive conclusions regarding this scenario, in particular, 
%To confirm the scenario, we need more detailed analysis using 
further \HI\ observations at higher angular resolutions. 

\begin{figure}
    \centering
    \includegraphics[width=\linewidth,clip]{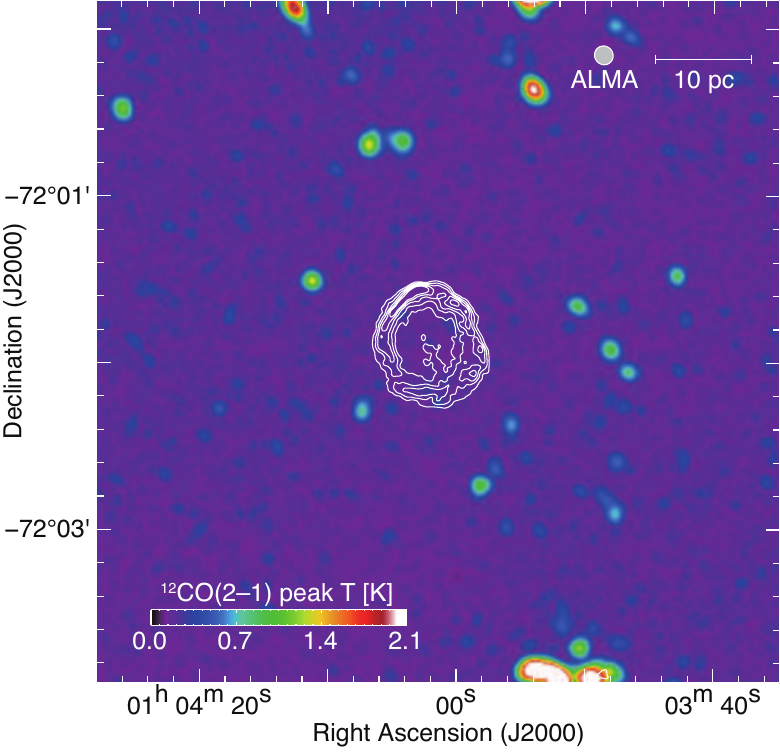}
    \caption{Peak temperature map of $^{12}$CO\,($J$~=~2--1) line emission towards E0102 obtained from ALMA.  Our ATCA 5500\,MHz image 
    is overlaid in contours, with levels as per Fig.~\ref{colored4}. %The contour lines are donated from 5500\,MHz image. The contour levels are as the same as Fig.~\ref{colored1}. 
    In the upper right-hand corner, the ALMA synthesised beam and a bar representing physical size on the sky is shown, while a colour bar at the bottom indicates CO gradients in K. }%The grey ellipse in the upper panand colour bar on the bottom left panel represent a synthesised beam and the gradients of CO in K, respectively.}
    \label{fig:CO}
\end{figure}

Multi--J~CO observations could also give better insight into the nature of E0102. Fig.~\ref{fig:CO} shows the peak temperature map of the  $^{12}$CO (J~=~2–1) emission line obtained with ALMA \citep{2021ApJ...922..171T}. Although several CO clumps are detected within the field of view shown in Fig.~\ref{fig:HI}, we find no dense CO clumps towards the radio continuum shell of E0102, suggesting it is interacting with only the \HI\ clouds. %This is consistent with the almost circular shape of the radio continuum shell without strong deformation.

%%%%%%%%%%%%%%%%%%%%%%%%%%%%%%%% Fig. 12 %%%%%%%%%%%%%%%%%%%%%%%%%%%%%%
%%%%%%%%%%%%%%%%%%%%%%%%%%%%%%%%%%%%%%%%%%%%%%%%%%%%%%%%%%%%%%%%

%%%%%%%%%%%%%%%%%%%%%%%%%%%%%%%% Fig. 13 %%%%%%%%%%%%%%%%%%%%%%%%%%%%%%
%%%%%%%%%%%%%%%%%%%%%%%%%%%%%%%%%%%%%%%%%%%%%%%%%%%%%%%%%%%%%%%%

%%%%%%%%%%%%%%%%%%%%%%%%%%%%%%%% Fig. 14 %%%%%%%%%%%%%%%%%%%%%%%%%%%%%%
%%%%%%%%%%%%%%%%%%%%%%%%%%%%%%%%%%%%%%%%%%%%%%%%%%%%%%%%%%%%%%%%

%jlp
%\section{Comparison with Similar Age core-collapse and Type\texorpdfstring{\,Ia}{Ia} Supernova Remnants} \label{com}

\section{Discussion} \label{com}

Our new radio continuum and polarisation study of E0102 reveal typical characteristics of a young SNR. We found a low integrated linear polarisation which indicates a high degree of turbulence. The total radio-continuum spectrum is rather steep which is also found in other young SNRs (see Table~\ref{tab3}) and we found areas of radial magnetic fields typically seen in younger remnants. In this section, we discuss our results in relation to E0102's environment and magnetic field comparing them to other young SNRs.
 
%We compare the physical characteristics of the E0102 SNR in the SMC, such as diameter, spectral index ($\alpha$), $P$, magnetic field strength, luminosity and surface brightness, with known and well studied young and middle-aged core-collapse and type\,Ia SNRs in both the Milky Way (MW) and LMC. These parameters are presented in Table~\ref{tab3}. 

%\Roland{R: Why have these SNRs been chosen to be compared to E0102? They cannot be the only young and middle aged SNRs. We need a justification for this sample.}\Rami{These SNRs are known and well studied.}
%We compare the physical characteristics of E0102 (diameter, spectral index ($\alpha$), $P$, magnetic field strength, luminosity, and surface brightness) with other young and middle age core-collapse and type\,Ia SNRs in MW and LMC as listed in Table~\ref{tab3}.

\subsection{The environment of E0102}

Our study of {\HI} line emission surrounding E0102 (Section~\ref{HI}) reveals the possibility of a stellar wind bubble; a closer look at the structure of this SNR's environment is therefore warranted. The measurements of \cite{2017A&A...602L...4V} indicate a pre-shock density of $n_0\approx 4\pm 1$\,cm$^{-3}$ ($n_0\approx 7.4\pm 1.5$~cm$^{-3}$) surrounding E0102, based on \FeXIV\ and \FeXI\ \OIII\ measurements. This density, however, appears high given typical densities of about 0.02\,cm$^{-3}$ \citep{1977ApJ...218..377W} in such wind-bubbles and would indicate that the SNR-shock has reached the ISM. Taking 7.4\,cm$^{-3}$ as the pre-shock density and using $r_{\rm w} = 56 n_0^{-0.3}$\,pc \citep{1984ApJ...278L.115M} to calculate the wind-bubble size, one finds $r_{\rm w}\geq30$\,pc, which is almost five times the radius of E0102. Further, the expansion rate derived by \cite{2017A&A...602L...4V} differs strongly from the rate derived by \cite{Xi_2019}. A possible explanation is that \cite{2017A&A...602L...4V} measured the downstream density and obtained an upstream density by assuming the canonical shock compression-ratio of $\approx4$. In general, the \FeXIV\ and \FeXI\ emissions are not associated with the outer-most regions of E0102. Hence, the emission can be an imprint of a previous interaction of the SNR-blastwave with a dense structure in the progenitor's wind, breaking the simple relation between up- and downstream density. Massive stars undergo different stages of evolution, and depending on their ZAMS-mass, winds get crushed by stronger winds of subsequent evolutionary stages. Shell-like features can emerge as a consequence and influence the following SNR evolution as shown for a massive progenitor with a crushed LBV-shell in \cite{2022A&A...661A.128D}. There is evidence that interaction with such a dense and asymmetric shell in the case of Cassiopeia A is responsible for the observed expansion dynamics of that remnant \citep{2022A&A...666A...2O}.

Given the possible presence of a dense shell in the progenitor's wind, it is likely that E0102 has already entered the Sedov-phase as suggested by \cite{Xi_2019}. However, the structure of wind-bubbles also featured a transition between the free-expanding wind, where $\rho\propto r^{-2}$, and a shocked wind with  roughly constant density. Further detailed modelling of the apparently complex environment of E0102 is beyond the scope of this paper.

\subsection{Comparison with other young SNRs}

Insight into our results for E0102 is compared to the broader SNR population by listing properties  (e.g. size, spectral index and polarisation) with other young SNRs ($\leq$1000\,yrs) in the MW and LMC (Table~\ref{tab3}).

\subsubsection{Diameter}
The diameter of SNRs from our selected sample varies from 0.4\,pc for the youngest (1987\,A; 36\,yr) to $\sim$23\,pc for the oldest (G266.2--1.2; 2400--5100\,yr). Although N\,132D and E0102 have similar ages ($\sim$2000\,yr), N\,132D is nearly twice the diameter. Perhaps this is related to the radio surface brightness at 1 GHz ($\Sigma_{\rm 1\,GHz}$ for N\,132D is 27.9$\times10^{-20}$\,W\,m$^{-2}$\,Hz$^{-1}$\,sr$^{-1}$ \citep{2017ApJS..230....2B} versus  9.7$\times10^{-20}$\,W\,m$^{-2}$\,Hz$^{-1}$\,sr$^{-1}$ for E0102, see Table~\ref{tab3}). %However, the size and the brightness might be the result of the same cause, but they do not cause each other. I don't know what this means
Typically one would expect a remnant with a higher ambient density to be brighter (and smaller), but a higher explosion energy could cause a SNR to be larger and brighter at the same age. The diameters for bright LMC SNRs N\,49 and N\,63A are $\sim$18\,pc which is comparable to the diameter of E0102. Selected type\,Ia (thermonuclear) SNRs (N\,103B, J0509--6731, LHG\,26, and Kepler) are slightly younger, which may account for their smaller diameters (7--8\,pc). The youngest SNRs from our sample (Cas\,A, G1.9+0.3 and Tycho) display smaller diameters, $\sim$5\,pc, as they have not yet expanded as far as their older counterparts. Various factors such as the composition of the ambient medium, explosion energy, and age all play an important role in the diameter of SNRs.

%{\bf MDF: need a lot of polishing here... difficult to read and understand the point! Why you didn't use 0509? not just environment but the difference in progenitor masses plays important role... Did you try to plot age vs diameter?}

%this is due to the difference in progenitor masses, the progenitor mass of N\,132D is $15\pm5$\,\(M_\odot\) \citep{2020ApJ...894..145S}

\subsubsection{Spectral Index} 
E0102 has a spectral index ($\alpha$) within the range of  SNRs listed in Table~\ref{tab3}, although there is a slight trend for older SNRs to have shallower values \citep{2017ApJS..230....2B,2022MNRAS.512..265F}. % with a slight preference for older SNR having shallower spectral indices. 
The spectral index for E0102, $\alpha = -0.61 \pm 0.01$, is similar to that of N\,132D, Kepler, and Tycho (see values and references in Table~\ref{tab3} which suggest a similar evolutionary phase). 
%{\bf From this particular information we can infer what with regards to our particular SNR?}
%with $\alpha = -0.65 \pm 0.04$ \citep{2017ApJS..230....2B}, Kepler with $\alpha = -0.64$ \citep{2014BASI...42...47G}, and Tycho with $\alpha = -0.6$ \citep{2014BASI...42...47G}. 
Other radio spectral indices vary from --0.81 in the youngest Galactic SNR (G1.9+0.3) to --0.3 in the oldest (G266.2--1.2) as expected
%the steepness of the radio spectral index is proportional to age
\citep[see e.g.][for review]{2014Ap&SS.354..541U}.

%{\bf MDF: perhaps you could try to plot age vs spectral index --- similar to Bozzetto+17. the problem is that your sample is small! }

%jlp marker

\subsubsection{Polarisation and Magnetic Field}
The mean fractional polarisation of SNRs varies as a function of age. Young SNR fractional 
polarisation can be as high as $\sim$20\,per\,cent, while older ones may have twice that value \citep{1990IAUS..140...81D} \footnote{For example, the Vela SNR has a fractional polarisation $>$40\,per\,cent \citep{,1995MNRAS.277.1435M}.}. 
Values of well-studied SNRs within the LMC, as shown in Table \ref{tab3}, vary from 3 to $\sim$26\,per\,cent .
%In Table~\ref{tab3}, we note that $P$ values of LMC SNRs are somewhat similar (3--8\,per\,cent) except J0509--6731 with $P$ value of $\sim$26\,per\,cent.
Since E0102 has fractional polarisations of $7\pm1$ and $12\pm2$ \,per\,cent (at 5500 and 9000\,MHz, respectively), we suggest its age and evolutionary phase (and possibly environmental conditions) are typical of this population.
%However, the $P$ value of E0102 is 7$\pm$1\,per\,cent. 

%
MW SNRs also show disparity in values of fractional polarisation ($P$). $P$ for the youngest Galactic SNR (G1.9+0.3) is very low (6\,per\,cent) compared with Tycho (20--30\,per\,cent) and SN\,1006 (17\,per\,cent). This diversity of $P$ may be due to the difference in ambient density and SNR age, as well as observational effects due to various radio telescope configurations. Fractional polarisation significantly changes with frequency because of Faraday rotation, which affects lower frequencies much more than higher frequencies. These issues make $P$ comparison difficult.
%However, our main results are more qualitative hence \textbf{are} not affected by this. 
%{\bf in fact is is higher than a couple of other sources too, so why this particular mention? Also, what does this then allow us to infer about this particular source}

%{\bf this last sentence does not contractdict of lead on from the previous one, so "while" does't work. also, what is the point in comparing the MW ones here? Maybe you want to write, `Interstingly CasA, a SNR in the MW, shows a high....' }

We estimate the line-of-sight magnetic field strength of E0102 is $\sim44$\,$\mu$G, which is higher than the value of the same age SNR N\,132D (Table~\ref{tab3}). However, the equipartition field of E0102 is $65\pm5$\,$\mu$G, similar to the N\,49 value of 91.2\,$\mu$G. Interestingly, Cas\,A, an oxygen-rich SNR in the MW, shows a high magnetic field strength value of 780\,$\mu$G \citep{2018A&A...612A.110A} due to the low value of $P$, indicative of a highly turbulent magnetic field \citep{alma9940430739605131}. Similar behaviour can be seen in other MW SNRs (see Table~\ref{tab3}). 

%{\bf so then why do you speficially mention CasA -is this source most likely to be the most similar to E0102?}

%from Roland
%Looking at the $RM$ map (Fig.~\ref{fig:RM_boxes}), it is very interesting that there is a $RM$ gradient on the bright shell in polarised intensity. It goes from about 400 to 800\,rad\,m$^2$. In this area the intrinsic magnetic field (Fig.~\ref{fig:MFD}) is tangential to the shell, indicating that that part of the SNR is already dominated by the ambient medium and therefore into the Sedov phase. At the edges of this area the magnetic field becomes gradually radial and spectral index significantly flatter. Most of the SNR has a radial magnetic field and where the magnetic field is radial the percentage polarisation is a lot lower. This all makes sense as young SNRs with radial magnetic fields have much more turbulence in their shells and therefore more depolarisation.

%The $RM$ gradient reflects the magnetic field configuration at the near-side of the SNR. Because that part is Faraday rotating more of the emission coming from the shell of the SNR. In this case, where the gradient is positive from the top to the left, the ambient magnetic field should be pointing from the left to the top. There is also a couple of areas at the bottom of the SNR where we can find a tangential magnetic field, most notably that bright blob in polarisation in the south.

%

\subsubsection{Surface Brightness and Luminosity} 

%\citet{1969AuJPh..22..107G} showed that the luminosity ($L$) of an SNR is inversely proportional to the fractional polarisation ($P$). %It is suggested that luminosity ($L$) is inversely proportional to $P$ \citep{1969AuJPh..22..107G}.
%We estimate the $L_{\rm 1\,GHz}$ of each SNR mentioned in this study (Table~\ref{tab3}) using the equation:
%\begin{equation}
%\label{eq3}
%L_{\rm 1\,GHz}=4\pi d^{2}S_{\rm 1\,GHz}
%\end{equation}

%\noindent where $L_{\rm 1\,\rm GHz}$ is the luminosity at 1\,GHz flux density (W\,Hz$^{-1}$), $d$ is the distance to the SNR in parsec, and $S_{\rm 1\,GHz}$ is radio flux at 1\,GHz in Jansky. 

%{\bf everywhere or just in this study? Clarify} 

The surface brightness ($\Sigma_{\rm 1\,GHz}$) for all SNRs used here for comparison shows, as expected, a variety of intensities.
%similar trend as $L_{\rm 1\,GHz}$. 
While, SNR~1987A displays a high $\Sigma_{\rm 1\,GHz}$ of $13700\times10^{-20}$\,W\,m$^{-2}$\,Hz$^{-1}$\,sr$^{-1}$ %is the highest which is expected as 1987\,A is 
as it is the youngest Local Group SNR, recently discovered intergalactic SNR J0624--6948 is barely detectable with a $\Sigma_{\rm 1\,GHz}$ of $0.0154\times10^{-20}$\,W\,m$^{-2}$\,Hz$^{-1}$\,sr$^{-1}$. We note  Cas\,A also has quite a high value $\Sigma_{\rm 1\,GHz}$ of $1145.27\times10^{-20}$\,W\,m$^{-2}$\,Hz$^{-1}$\,sr$^{-1}$ (Table~\ref{tab3}). 

%{\bf what does this mean? Why is this information significant? How does this information help you to understand your source E0102?}

Comparing a previous $\Sigma$--D diagram \citep[][their fig.~3]{2022PASP..134f1001U,2020NatAs...4..910U,2018ApJ...852...84P} and our values, D~=~$\sim$13\,pc and $\Sigma_{\rm 1\,GHz} =9.17\times10^{-20}$\,W\,m$^{-2}$\,Hz$^{-1}$\,sr$^{-1}$ for E0102 (see Table~\ref{tab3}), E0102 is in the middle of their $\Sigma$--D diagram. This is consistent with a Sedov phase having a dense ambient density of $n_{\rm H}=0.5$\,cm$^{-3}$.

Finally, we note that E0102 has a $L_{\rm 1\,GHz}$\footnote{We estimate the luminosity at 1\,GHz using:
$L_{\rm 1\,GHz}=4\pi d^{2}S_{\rm 1\,GHz}$, where $d$ is distance and $S_{\rm 1\,GHz}$ is flux density at 1 GHz.} of 1.4$\times10^{17}$\,W\,Hz$^{-1}$, which is  less than the values of some LMC SNRs (N\,49 and N\,63A), while similar age N\,132D shows a much higher value (15.7$\times10^{17}$\,W\,Hz$^{-1}$). The MW SNRs have less $L_{\rm 1\,GHz}$ than E0102, apart from Cas\,A which shows a high value of 36.1$\times10^{17}$\,W\,Hz$^{-1}$. %This makes E0102 one of the most luminous young SNRs compared to those within %both, MCs and-if you are saying the luminosity is less than several LMC SNRs, how can you say it is the most luminous in the MC??
%the MW.
Therefore, E0102 has a $L_{\rm 1\,GHz}$ less than many young LMC SNRs but greater than most young SNRs in the MW.

%%%%%%%%%%%%%%%%%%%%%%%%%%%%%%% table 3 %%%%%%%%%%%%%%%%%%%%%%%%%%%%%%%%%%
%\newpage
%\begin{landscape}
    \begin{table*}
    \centering
\vspace{5cm}

	\caption{Comparison of E0102 with young SNRs ($\leq$1000\,yrs) in the MW and LMC. SN type is denoted by CC -- Core-Collapse, and TN -- Thermonuclear (type\,Ia). Magnetic field strength for the molecular cloud associated with the SNR is marked with an asterisk ($^{*}$) and magnetic field strength estimation using equipartition is marked with an double asterisk ($^{**}$). }
	\label{tab3}
	\begin{tabular}{@{}lllcccccccccc@{}}
		\hline
Name                &Host       &   SN     &     Age           & Diameter          &    Spectral Index    & Avg. $P$         &$P_{\nu}$    & Mag. Field       & $\Sigma_{\rm 1\,GHz}(\times10^{-20})$      \\
                   &Galaxy     &   Type   &     (yr)          &  (pc)             &      ($\alpha$)        &   (\%)           &   (MHz)     & ($\mu$G)         & (W\,m$^{-2}$\,Hz$^{-1}$\,sr$^{-1}$)       \\  
	\hline
E0102               & SMC       &    CC    &    $\sim1738^{1}$ &  $\sim13$         &    $-0.61\pm0.01$    &     $7\pm1/12\pm2$      & $5500/9000$      & $65\pm5^{**}$            &         $9.17$                             \\
\hline
J0624--6948         &LMC        & TN$^{2}$ &     $2200^{2}$    &    $47.5^{2}$     &    $-0.54\pm0.08^{2}$&    $\le 9^2$           &   $5500^2$      &   ---             &           $0.0154^{2}$                     \\
N\,49               &LMC        &    CC    &   $\sim4800^{3}$  & $18.2^{4}$        & $-0.56\pm0.03^{4}$   &   $4.8^{5}$      & $8640^{5}$ & $91.2^{**}$       &          $16^{4}$                          \\
N\,132D             &LMC        &    CC    & $\sim2500^{6}$    & $24.5^{4}$        &$-0.65\pm0.04^{4}$    & $4^{7}$          & $4790^{7}$ & $2.1^{7}$         &          $27.9^{4}$                       \\
N\,63A              &LMC        &    CC    & $3500\pm1500^{8}$ & $17.8^{4}$        &$-0.740\pm0.002^{8}$  &   $5\pm1^{9}$    & $5500^{9}$ & $80-180^{9*}$     &          $18.7^{4}$                       \\
%N158A  &LMC        & $1100 \pm340^{5}$  &$-0.63\pm0.03^{6}$  &  $^{7}$         &  \\
1987\,A             & LMC       &    CC    & $34$              &  $0.4^{4}$        &$-0.8^{10}$            & $2.7\pm0.2^{11}$ &$22000^{11}$ & $\sim28^{11}$    &         $13700^{4}$                      \\
N\,103B             & LMC       &    TN    & $380-860^{12}$    & $6.8^{4}$         &$\sim -0.75^{13}$     &  $8\pm1^{13}$    &$5500^{13}$  & $0.03-82.1^{13}$ &       $60^{13}$                           \\
J0509--6731         &LMC        &    TN    & $\sim310^{14}$    &  $7.4^{15}$       & $-0.73^{16}$         & $\sim26^{16}$    & $5500^{15}$ & $\sim128^{15**}$ &           $5.72^{4}$                      \\
LHG\,26             &LMC        &    TN    & $\sim600^{16}$    &  $8.3^{16}$       & $\sim-0.54^{17}$     & $\sim8^{17}$     & $5500^{17}$ & $\sim171^{17**}$ &           $6.09^{4}$                      \\
Cas\,A              & MW        &    CC    & $325^{18}$        &$\sim5^{19}$       &$-0.77^{20}$          &  $4.5^{21}$      &$19000^{21}$ & $780^{19}$       &          $1145.27^{22}$                   \\
G1.9+0.3            & MW        &CC$^{23}$ & $\sim120^{24}$    &  $4^{25}$         &$-0.81\pm0.02^{23}$   & $6^{26}$         &$5500^{26}$  & $\sim273^{26**}$ &        $6.7^{23}$                             \\
G266.2--1.2         & MW        &    CC    &  $2400-5100^{27}$ &$23.60\pm0.06^{28}$&$\sim-0.3^{28}$       &  ---             &   ---       & $6-10^{29}$      &       $0.05^{22}$                         \\
Tycho               & MW        &    TN    &    $451^{30}$     &  $3.5-7.2^{30}$   &   $-0.58^{31}$       & $20-30^{32}$     &$4872^{32}$  & $50-400^{33}$    &         $13.17^{22}$                      \\
SN1006              &MW         &    TN    & $1017^{34}$       & $\sim19^{34}$     & $-0.6^{31}$          & $\sim17^{35}$    & $1400^{35}$ & $30-40^{36}$     &        $0.32^{22}$                        \\
Kepler              &MW         &    TN    & $419^{37}$        &  $8.18^{37}$      & $-0.64^{31}$         & $\sim6^{38}$     & $4835^{38}$ & $\sim414^{39**}$ &         $31.78^{22}$                      \\
G11.2$-$0.3         &MW         &    CC    & $1636^{40}\dag$       &  $6.0$            & $-0.56^{41}$         & $\sim2^{42}$     & $32000^{42}$& ---              &                $20.7^{31}$                     \\
G29.7--0.3          &MW         &    CC    &  $\sim480^{43}$   &   $5.0^{43}$      &        $-0.63\pm 0.02^{44}$           &     $\le1^{45}$          &     $2695^{45}$     &        ---       &            $16.7^{31}$                            \\
%G353.6--0.7        & MW        &    CC    & $\sim27000^{37}$    & $3.2^{38}$ &$\sim28^{37}$   & $-1.1\le\alpha\le-0.15^{39}$& ---    &   ---      &  $25^{40}$&$0.03$   & $0.04^{16}$ &$2.5^{17}$ \\
    
	\hline
	\end{tabular}

	\begin{tablenotes}
      \footnotesize
      \item 
      \item References: (1) \cite{2021ApJ}, (2) \citet{2022MNRAS.512..265F}, (3) \citet{2012ApJ...748..117P}, (4) \citet{2017ApJS..230....2B}, (5) \citet{1998AJ....115.1057D}, (6) \citet{2020ApJ...894...73L}, (7) \citet{1995AJ....109..200D}, (8) \citet{2003ApJ...583..260W}, (9) \citet{2019ApJ...873...40S}, (10) \citet{2013ApJ...767...98Z}, (11) \citet{2018ApJ...861L...9Z}, (12) \citet{2005Natur.438.1132R}, (13) \citet{2019Ap&SS.364..204A}, (14) \citet{2018MNRAS.479.1800R}, (15) \citet{2014MNRAS.440.3220B}, (16) \citet{2006ApJ...642L.141B}, (17) \citet{2012SerAJ.185...25B}, (18) \citet{2006ApJ...645..283F}, (19) \citet{2018A&A...612A.110A}, (20) \citet{2014ApJ...785....7D}, (21) \citet{1968ApJ...151...53M}, (22) \citet{2019SerAJ.199...23S}, (23) \citet{2020MNRAS.492.2606L}, (24) \citet{2017MNRAS.468.1616P}, (25) \citet{2008ApJ...680L..41R}, (26) \citet{2014SerAJ.189...41D}, (27) \citet{2015ApJ...798...82A}, (28) \citet{2018ApJ...866...76M}, (29) \citet{2013A&A...551A.132K}, (30) \citet{2000ApJ...545L..53H}, (31) \citet{2014BASI...42...47G}, (32) \citet{1991AJ....101.2151D}, (33) \citet{2015ApJ...812..101T}, (34) \citet{2003ApJ...585..324W}, (35) \citet{2013AJ....145..104R}, (36) \citet{2006ESASP.604..319V}, (37) \citet{2012ApJ...756....6P}, (38) \citet{2002ApJ...580..914D}, (39) \citet{2012ApJ...746...79A}, (40) \citet{1977hisu.book.....C}, (41) \citet{2002ApJ...572..202T}, (42) \citet{2001A&A...372..627K}, (43) \citet{2018ApJ...856..133R}, (44) \citet{2011A&A...536A..83S}, (45) \citet{1999A&A...350..447D}. \\
      \dag Note that \citet{2018ApJ...865L...6Z} argues that SNR G7.7--3.7 is more likely SN from 386CE.
      %(37) \citet{2008ApJ...679L..85T}, (38) \citet{2010ApJ...712..790T}, (39) \citet{2017MNRAS.467..155N}, (40) \citet{2011A&A...531A..81H}.
      \end{tablenotes}
      \end{table*}

\section{Conclusions} 
 \label{con}
% We present new high-resolution and sensitive ATCA images for SMC SNR E0102 at 5500 and 9000\,MHz, which exhibit a ring-like morphology in both images. The radius of this remnant varies from \textbf{6.89\,pc} towards the south-west to \textbf{6.06\,pc} towards the south-east, \textbf{6.06\,pc} towards the north-west, and \textbf{5.82\,pc towards} north-east. 

% The 5500\,MHz image shows a bridge-like structure seen for the first time in a radio image. This structure is consistent with both optical and X-ray images. In the 9000\,MHz image we detect a central region, this region might be a PWN. However, we doubt this possibility since this central region is unpolarised. 

% We estimated the $P$ values for this SNR of $7\pm1$ and $12\pm2$\,per\,cent for 5500 and 9000\,MHz, respectively. The polarisation is concentrated on the north-east part of the remnant. We estimated the mean value of the radio spectral index for the whole remnant of $\bm{\alpha=-0.61\pm0.01}$~--~ in line with values for similar age SNRs in the LMC \textbf{and MW}. We calculated the line of sight magnetic field strength for this SNR of 44.2\,$\mu$G, and we also used the equipartition formula to estimate the magnetic field strength of \textbf{65.1/147.8\,$\mu$G} for $\kappa=0$/$\kappa\neq0$. We detected an \HI\ cloud towards this remnant at the velocity range of $\sim$160--180\,km\,s$^{-1}$ and a cavity-like structure at the velocity of 163.7--167.6\,km\,s$^{-1}$. We did not detect CO emission towards E0102.
SNR E0102 is a young SNR located in the SMC with properties including spectral index and luminosity consistent with young SNRs in the MW and
LMC. Our new radio continuum images of E0102 show a ring morphology and bridge-like structure seen in both optical and X-ray images. Our new images also show a central feature that lacks any significant polarisation towards it suggesting that the origin is not a PWN. We find that the remnant's radio emission extends beyond its X-ray emission tracing a forward and reverse shock with [Fe \textsc{XIV}] emission sandwiched between possibly representing denser ionized clumps. The remnant shows polarised regions in its shell and it was possible to obtain a rotation measure and estimate its magnetic field. Comparison with theoretical intrinsic magnetic field estimates and location on a $\Sigma-D$ diagram of comparable SNRs are constant with it in or entering its Sadov phase with a high ambient density.

\section*{Acknowledgements}
The Australia Telescope Compact Array (ATCA) and Australian SKA Pathﬁnder (ASKAP) are part of the Australia Telescope National Facility which is managed by CSIRO. 
%Operation of the ASKAP is funded by the Australian Government with support from the National Collaborative Research Infrastructure Strategy. The ASKAP uses the resources of the Pawsey Supercomputing Centre. Establishment of the ASKAP, the Murchison Radio-astronomy Observatory, and the Pawsey Supercomputing Centre are initiatives of the Australian Government, with support from the Government of Western Australia and the Science and Industry Endowment Fund. We acknowledge the Wajarri Yamatji people as the traditional owners of the Observatory site. 
This paper makes use of the following ALMA data: ADS/JAO.ALMA\#2017.A.00054.S and ADS/JAO.ALMA\#2018.1.01031.S. ALMA is a partnership of ESO (representing its member states), NSF (USA) and NINS (Japan), together with NRC (Canada), MOST and ASIAA (Taiwan), and KASI (Republic of Korea), in cooperation with the Republic of Chile. The Joint ALMA Observatory is operated by ESO, AUI/NRAO and NAOJ. This research has made use of the software provided by the Chandra X-ray Centre (CXC) in the application packages CIAO (v 4.12). This work was supported by JSPS KAKENHI Grant Numbers JP19H05075 (H. Sano) and JP19K14758 (H. Sano). S.D. is the recipient of an Australian Research Council Discovery Early Career Award (DE210101738) funded by the Australian Government. A.J.R. acknowledges financial support from the Australian Research Council under award number FT170100243. M.S. and J.K. acknowledge support from the Deutsche Forschungsgemeinschaft through the grants SA 2131/13-1, SA 2131/14-1, and SA 2131/15-1. C.C. acknowledges support from the Studienstiftung des deutschen Volkes. H.S. was supported by Grants-in-Aid for Scientific Research 
(KAKENHI) of Japan Society for the Promotion of Science 
(JSPS; grant Nos. JP21H01136, and JP19K14758).
K.T. was supported by NAOJ ALMA Scientific Research grant No.2022-22B and Grants-in-Aid for Scientific Research (KAKENHI) of Japan Society for the Promotion of Science (JSPS; grant Nos. JP21H00049, and JP21K13962). D.U. acknowledges the Ministry of Education, Science and Technological Development of the Republic of Serbia through the contract No. 451-03-68/2022-14/200104, and for the support through the joint project of the Serbian Academy of Sciences and Arts and Bulgarian Academy of Sciences on the detection of extragalactic SNRs and \HII\ regions. R.B. acknowledges funding from the Irish Research Council under the Government of Ireland Postdoctoral Fellowship program. Based on observations made with ESO Telescopes at the La Silla Paranal Observatory under programme ID 297.D-5058[A].

\section*{Data Availability}

The data that support the plots/images within this paper and other findings of this study are available from the corresponding author upon reasonable request. The ASKAP data used in this article are available through the CSIRO ASKAP Science Data Archive (CASDA) and ATCA data via the Australia Telescope Online
Archive (ATOA).

%%%%%%%%%%%%%%%%%%%%%%%%%%%%%%%%%%%%%%%%%%%%%%%%%%



\bibliographystyle{mnras.bst}
\bibliography{example} % if your bibtex file is called example.bib

\section*{}
Please note: Oxford University Press is not responsible for the content or functionality of any supporting materials supplied by the authors. Any queries (other than missing material) should be directed to the corresponding author for the article.

\section*{}
{\it 
% List of institutions
$^{1}$Western Sydney University, Locked Bag 1797, Penrith, NSW, 2751, Australia\\
$^{2}$CSIRO Astronomy and Space Sciences, Australia Telescope National Facility, PO Box 76, Epping, NSW 1710, Australia\\
$^{3}$Faculty of Engineering, Gifu University, 1-1 Yanagido, Gifu 501-1193, Japan\\
$^{4}$National Astronomical Observatory of Japan, Mitaka, Tokyo 181-8588, Japan\\
$^{5}$Dominion Radio Astrophysical Observatory, Herzberg Astronomy and Astrophysics, National Research Council Canada, PO Box 248, Penticton BC V2A 6J9, Canada\\
$^{6}$Dublin Institute for Advanced Studies, Astronomy \& Astrophysics Section, 31 Fitzwilliam Place, D02 XF86 Dublin 2, Ireland \\
$^{7}$Dr. Karl Remeis Observatory, Erlangen Centre for Astroparticle Physics, Friedrich-Alexander University Erlangen-N\"{u}rnberg, Sternwartstr. 7, 96049 Bamberg, Germany\\
$^{8}$Max-Planck-Institut f\"{u}r extraterrestrische Physik, Gie{\ss}enbachstra{\ss}e 1, 85748 Garching, Germany\\
$^{9}$School of Cosmic Physics, Dublin Institute for Advanced Studies, 31 Fitzwillam Place, Dublin 2, Ireland\\
$^{10}$Department of Physics and Astronomy, University of Calgary, University of Calgary, Calgary, Alberta, T2N 1N4, Canada\\
$^{11}$Observatoire Astronomique de Strasbourg, Universit\'e de Strasbourg, CNRS, 11 rue de l'Universit\'e, F-67000 Strasbourg, France\\
$^{12}$European Southern Observatory, Alonso de C\'ordova 3107, Vitacura 763 0355, Santiago, Chile\\
$^{13}$Joint ALMA Observatory, Alonso de C\'ordova 3107, Vitacura 763 0355, Santiago, Chile\\
$^{14}$School of Physical Sciences, The University of Adelaide, Adelaide 5005, Australia\\
$^{15}$School of Science, University of New South Wales, Australian Defence Force Academy, Canberra, ACT 2600, Australia\\
$^{16}$Department of Earth and Planetary Sciences, Faculty of Sciences, Kyushu University, Nishi-ku, Fukuoka 819-0395, Japan\\
$^{17}$National Astronomical Observatory of Japan, National Institutes of Natural Sciences, 2-21-1 Osawa, Mitaka, Tokyo 181-8588, Japan\\
$^{18}$Department of Physics, Graduate School of Science, Osaka Metropolitan University, 1-1 Gakuen-cho, Naka-ku, Sakai, Osaka 599-8531, Japan\\
$^{19}$Department of Astronomy, Faculty of Mathematics, University of Belgrade, Studentski trg 16, 11000 Belgrade, Serbia\\
$^{20}$Isaac Newton Institute of Chile, Yugoslavia Branch\\
$^{21}$Lennard-Jones Laboratories, Keele University, Staffordshire ST5 5BG, UK\\
$^{22}$Federal Office of Meteorology and Climatology - MeteoSwiss, Chemin de l’A\'erologie 1, 1530 Payerne, Switzerland}

\bsp	% typesetting comment
\label{lastpage}
\end{document}